\documentclass[a4paper,11pt]{article}
\pdfoutput=1
\usepackage{setup}
\usepackage{jheppub}

\usepackage{cite}
\usepackage{ifthen}
\usepackage{subcaption}
\usepackage{siunitx}
\usepackage{enumitem}
\usepackage{booktabs}

\DeclareSIUnit\fb{\femto\barn}

\renewcommand{\tabcolsep}{1em}

\makeatletter
\g@addto@macro\bfseries\boldmath
\makeatother

\definecolor{lightgray}{HTML}{A6A39A}
\definecolor{darkgray}{HTML}{504E48}
\definecolor{silver}{HTML}{E0DFDE}
\definecolor{brown}{HTML}{5F4541}
\definecolor{beige}{HTML}{DCCCAC}
\definecolor{green}{HTML}{345F53}
\definecolor{yellow}{HTML}{F6B65A}
\definecolor{blue}{HTML}{568BCF}
\definecolor{red}{HTML}{AE1932}
\definecolor{orange}{HTML}{D16F15}

\makeatletter
\newcommand{\myitem}[1]{%
	\item[#1]\protected@edef\@currentlabel{#1}%
}
\makeatother

\preprint{{\raggedleft%
NIKHEF 2021-026,
BONN-TH-2021-09, \\
IPPP/21/26,
ZU-TH 49/21,
CERN-TH-2021-159 \\
}}

\title{$V\PH+\jet$ production in hadron-hadron collisions up to order $\alphas^3$ in perturbative QCD}

\author[a,b]{R.~Gauld,}
\author[c,d]{A.~Gehrmann--De Ridder,}
\author[e,f]{E.~W.~N.~Glover,}
\author[g]{A.~Huss,}
\author[c]{I.~Majer.}

\affiliation[a]{Nikhef Theory Group, Science Park 105, 1098 XG Amsterdam, The Netherlands}
\affiliation[b]{Bethe Center for Theoretical Physics \& Physikalisches Institut der Universit\"at Bonn}
\affiliation[c]{Institute for Theoretical Physics, ETH, CH-8093 Z\"urich, Switzerland}
\affiliation[d]{Department of Physics, University of Z\"urich, CH-8057 Z\"urich, Switzerland}
\affiliation[e]{Institute for Particle Physics Phenomenology, Durham University,  Durham DH1 3LE, UK}
\affiliation[f]{Physics Department, Durham University,  Durham DH1 3LE, UK}
\affiliation[g]{Theoretical Physics Department, CERN, CH-1211 Geneva 23, Switzerland}

\emailAdd{rgauld@uni-bonn.de}
\emailAdd{gehra@phys.ethz.ch}
\emailAdd{e.w.n.glover@durham.ac.uk}
\emailAdd{alexander.huss@cern.ch}
\emailAdd{majeri@phys.ethz.ch}

\abstract{
We present precise predictions for the hadronic production of an on-shell Higgs boson in association with a leptonically decaying gauge boson and a jet up to order $\alphas^3$.
We include the complete set of NNLO QCD corrections to both charged- and neutral-current Drell-Yan type contributions, as well as the previously known leading heavy quark loop induced contributions which involve a direct Higgs--quark coupling.
As an application, we study a range of differential observables in proton--proton collisions at $\sqrt{s} = \SI{13}{\TeV}$ for both the charged- and neutral-current production modes.
For each Higgs production process, we assess the improvement in the theoretical uncertainty for both the exclusive ($n_\jet = 1$) and inclusive ($n_\jet \geq 1$) jet categories.
We find that the inclusion of the NNLO corrections to the Drell-Yan type contributions is essential in stabilising the predictions and in reducing the theoretical uncertainty for both inclusive and exclusive jet production for all three modes.
This is particularly true in the kinematical regimes associated with low to medium values of the transverse momentum of the produced vector boson and where the differential cross sections are the largest.
For the neutral-current process, we find that the heavy quark loop induced contributions have their largest phenomenological impact
(an increase in the size of the NNLO corrections, a distortion of the distribution shape and an enlargement of the left over remaining uncertainties) in kinematical regions associated to large values of $p_{T,Z}$ (typically above 150~GeV) where the cross sections are smaller.
}

\input{fig}

\begin{document}

\maketitle
\flushbottom


\section{Introduction}
\label{intro}

Since the discovery of the Higgs boson~\cite{Aad:2c012tfa,Chatrchyan:2012xdj}, the continued measurement of the properties and interactions of the Higgs boson with other fundamental particles has been a main goal of the LHC physics programme.
In particular, this concerns the coupling strengths of the Higgs boson to known Standard Model (SM) particles such as gauge bosons and fermions through the study of differential observables related to various Higgs boson production and decays channels.

A key channel is the associated production of a Higgs boson with an electroweak gauge boson, referred to as the Higgs-Strahlung or $V\PH$ channel ($V=\PWpm$ or $\PZ$).
The presence of a leptonically decaying gauge boson in the final state offers a handle to efficiently reject the contribution of multi-jet backgrounds.
This is critical for the measurement of the hadronic decays of the Higgs boson, which are otherwise overwhelmed by such background processes.
An important example is the decay of the Higgs boson into a pair of $\Pqb$-quarks, which in the SM occurs with an expected branching fraction of $\approx 58\%$.
Measurements of this decay channel (which are uniquely accessible in the $V\PH$ channel) are essential to directly probe the interaction strength of the Higgs boson with quarks, and also provide important constraints on the total decay width of the Higgs boson.

Experimentally, the $\PH\to\Pqb\Paqb$ decay was observed by the ATLAS~\cite{Aaboud:2018zhk} and CMS~\cite{Sirunyan:2018kst} collaborations in 2017, precisely through the $V\PH$ production mode with a significance of \numlist{5.6;5.3} standard deviations for CMS and ATLAS respectively.
In addition, first differential measurements based on simplified template cross sections as a function of the transverse momentum of the vector boson have been reported in ref.~\cite{Aaboud:2019nan} and updated including the full Run II data set in refs.~\cite{Aad:2020jym,Aad:2020eiv}. The latter measurements indicate that the observed production rates are presently consistent with SM expectations within experimental uncertainties of the order of 20 percent.

These measurements are performed by categorising the observed event rates according to the number of charged leptons (\Pl) and hadronic jets. Such selections are necessary to improve the experimental sensitivity to the signal processes.
In all channels, events are required to have exactly two \Pqb-tagged jets, which form the Higgs boson candidate. The 0, 1 and 2 lepton channels respectively provide access to the $V\PH$ decay processes: $\PZ\PH \to \nu \nu \Pqb\Paqb$, $\PW\PH\to\Pl\Pgnl\Pqb\Paqb$ and $\PZ\PH\to\Plp\Plm\Pqb\Paqb$.
Events are then further categorised based on the presence of hadronic jets which are not \Pqb-tagged (i.e.\ in addition to those two which form the Higgs candidate).
Examples are: $n_\jet = 0$, no additional hadronic jets; $n_\jet = 1$, exactly one additional hadronic jet; $n_\jet \geq 1$, at least one additional hadronic jet.
Notably, almost half of the selected events in an experimental analysis \cite{Aad:2020jym, Aad:2020eiv} are contained in the $n_\jet = 1$ or $n_\jet \geq 1$ categories. These latter categories will be the focus of this work.

As the experimental analyses rely on the theoretical knowledge of the $V\PH\,(+\jet)$ production and decay modes to interpret the data, it is of crucial importance to have the most accurate theoretical predictions available for differential observables in the selected event categories.
Depending on the event category, various levels of theoretical precision are currently available:
At fixed-order accuracy, NNLO QCD corrections for the $V\PH$ production mode have been available for some time~\cite{Brein:2003wg,Brein:2012ne,Ferrera:2011bk,Ferrera:2013yga,Ferrera:2014lca}.
More recently, calculations which also include decay corrections of the Higgs boson up to NLO~\cite{Campbell:2016jau} and NNLO QCD accuracy in the five-flavour~\cite{Ferrera:2017zex,Caola:2017xuq,Gauld:2019yng}\footnote{Those calculations necessarily rely on the use of a flavoured jet algorithm~\cite{Banfi:2006hf}.} and four-flavour schemes~\cite{Behring:2020uzq,Bizon:2021rww} have been presented.
Other NNLO QCD calculations including an interface to a parton shower event generator have been presented in~\cite{Astill:2016hpa,Alioli:2019qzz} and~\cite{Astill:2018ivh}, where the latter includes decay corrections.
The phenomenological impact of heavy quark loop contributions has been studied in~\cite{Brein:2011vx,Altenkamp:2012sx,Hespel:2015zea,Goncalves:2015mfa,Wang:2021rxu}.
For the $V\PH+\jet$ production mode, fixed-order NLO QCD computations for $V\PH$ and $V\PH+\jet$ production have been merged in the context of parton showers to provide full NLO+PS simulations for these (on-shell) Higgs production modes~\cite{Luisoni:2013kna,Astill:2018ivh}. The corresponding results have been further complemented with electroweak corrections in~\cite{Granata:2017iod}.
More recently, predictions including NNLO QCD corrections for differential observables related to the $W^+\PH+\jet$ production mode were presented (by us) in ref.~\cite{Gauld:2020ced}.

It is the purpose of this paper to provide precise predictions for observables related to all three $V\PH+\jet$ production modes with $V=\PZ,\,\PWp,\,\PWm$ where the vector boson decays leptonically and the Higgs boson is produced on its mass shell.
In particular, we increase the perturbative accuracy of the Drell-Yan-type component of the cross-section (which is numerically dominant) to NNLO QCD accuracy for all production modes.
The extension of these predictions to higher orders, including corrections to Drell-Yan type and heavy-quark loop-induced contributions up to the third order in $\alphas$, provides a new level of theoretical precision for these $V\PH+\jet$ production modes. 

This improvement will be vital for future interpretations of experimental measurements, as the current knowledge of the theoretical modelling of the signal process (currently limited to NLO accuracy in $n_\jet > 0$ categories) constitutes one of the dominant systematic uncertainties in signal extractions of these Higgs-Strahlung processes~\cite{ATLAS:2020fcp}.

The structure of the paper is as follows:
In section~\ref{details} we summarise the general structure of the calculation, and provide details of the various ingredients which enter the computations for the three distinct $V\PH+\jet$ production modes, up to order $\alphas^3$.
Our numerical results are presented in sections~\ref{fiducial} and \ref{distributions} for the case of $\Pp\Pp$ collisions at $\SI{13}{\TeV}$, where both inclusive and exclusive jet categories are considered and compared.
Specifically, we provide predictions for fiducial cross sections in section~\ref{fiducial} and differential distributions in section~\ref{distributions}, and discuss the phenomenological importance of including higher-order corrections in both cases.
Section \ref{conclusions} presents a summary of the results obtained together with an outlook for further studies.
Some of the results associated to $\PWp\PH+\jet$ production have been previously presented in ref.~\cite{Gauld:2020ced} and will be included here to facilitate a direct comparison with the other production modes, $\PWm\PH+\jet$ and  $\PZ\PH+\jet$, which are presented here for the first time.


\section{Details of the calculation}
\label{details}

In this work, differential observables for the production of a Higgs boson in association with a weak gauge boson and at least one resolved jet are computed up to and including order $\alphas^3$ contributions in perturbative QCD.
More specifically, we consider the two charged-current processes
\begin{subequations}
\label{eq:procs}
\begin{align}
   \label{eq:CC}
   \Pp \Pp &\to  \PH\PWpm + \jet \to \PH + \Plpm \Pgnl + \jet
\intertext{and the neutral-current process,}
   \label{eq:NC}
   \Pp \Pp &\to  \PH\PZ + \jet \to \PH + \Plm \Plp + \jet \,,
\end{align}
\end{subequations}
where the Higgs boson is produced on its mass shell and the weak boson decays leptonically.
The calculation is carried out within the \nnlojet framework~\cite{Ridder:2015dxa} which is a fixed-order parton-level event generator that employs the antenna subtraction formalism~\cite{GehrmannDeRidder:2005cm,GehrmannDeRidder:2005aw,GehrmannDeRidder:2005hi,Daleo:2006xa,Daleo:2009yj,Boughezal:2010mc,Gehrmann:2011wi,GehrmannDeRidder:2012ja,Currie:2013vh} for the treatment of infrared divergences in order to retain the full kinematic information on the final state.

\begin{figure}[t]
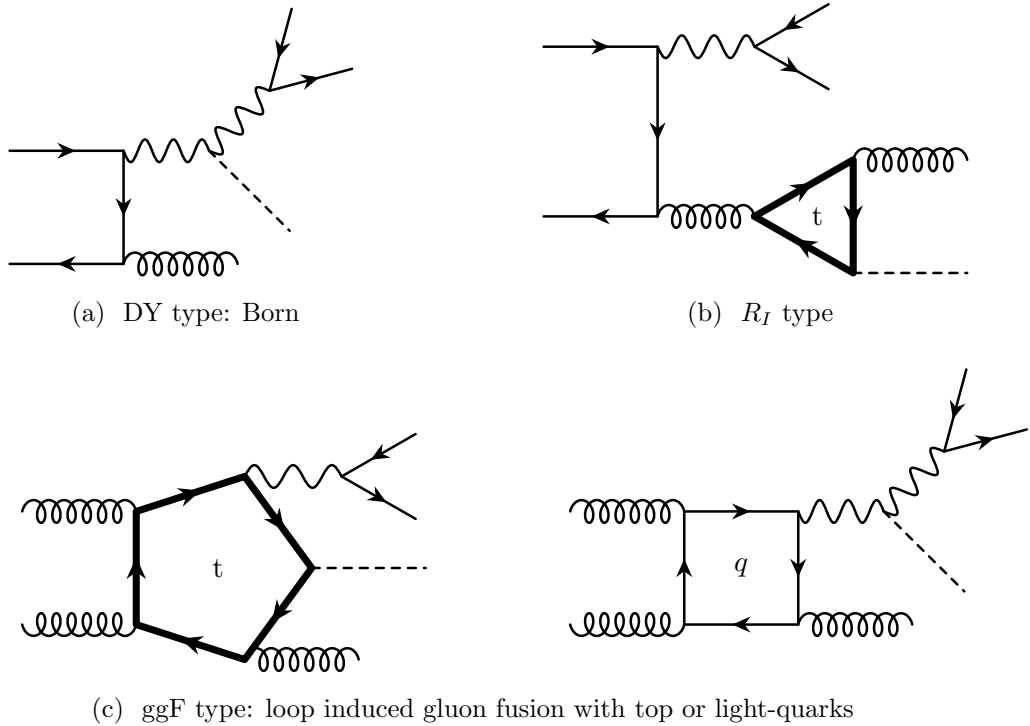

   \centering
   \begin{subfigure}[b]{.5\textwidth}
      \centering
      \BornVHJ{1.5cm}
      \caption{\label{fig:diag_born} \DY type: Born}
   \end{subfigure}%
   \begin{subfigure}[b]{.5\textwidth}
      \centering
      \TopRI{1.5cm}
      \caption{\label{fig:diag_ri} \RI type}
   \end{subfigure}
   \\\bigskip
   \begin{subfigure}[b]{\textwidth}
      \hfill
      \TopZHJI{1.5cm}
      \hfill
      \TopZHJII{1.5cm}
      \hfill
      \caption{\label{fig:diag_gg_top} \GGF type: loop induced gluon fusion with top or light-quarks}
   \end{subfigure}%
  \caption{Representative diagrams for the Drell-Yan and heavy-quark loop induced contributions to $V\PH+\jet$ production that enter the cross section at specific order in $\alphas$:
  (a) at order $\alphas$ for the Drell-Yan process, denoted as \DY-type;
  (b) at order $\alphas^2$ for the top-loop induced processes, denoted as \RI-type;
  and (c) at order $\alphas^3$ for the one-loop gluon--gluon initiated process $\Pg\Pg\to\PH\PZ\Pg$, denoted as \GGF-type.
  Details on the individual contributions given in the main text.}
  \label{fig:diagrams}
\end{figure}

At leading-order, the processes in Eq.~\eqref{eq:procs} give rise to cross-section contributions that are proportional to $\alphas$.
These are given by Higgs-Strahlung diagrams, where a Higgs boson is emitted from a massive gauge boson of an underlying Drell-Yan type $V+\jet$ process.
An example of Born-level diagram is depicted in figure~\ref{fig:diag_born}.
In the present calculation, full NNLO QCD corrections to the Drell-Yan type contributions (denoted as DY) are considered, i.e.\ up to order $\alphas^3$ for all three Higgs production modes.

Starting from order $\alphas^2$, an additional class of subprocesses contributes in which the Higgs boson couples to a virtual heavy quark loop.
In our calculation, we consider leading-order contributions from such heavy quark loop induced contributions up to order $\alphas^3$:
For both charged- and neutral-current processes this includes the top-quark loop corrections at order $\alphas^2y_\Pqt$, denoted as the \RI-type contributions and depicted in figure~\ref{fig:diag_ri}.
In the case of the neutral-current process, additional loop-induced contributions related to the gluon-fusion process $\Pg\Pg\to\PH\PZ\Pg$ must be considered. Those are proportional to $\alphas^3$ and illustrated in figure~\ref{fig:diag_gg_top}. In the following those are denoted as \GGF-type.

The various ingredients necessary for the computation of the $\PWp\PH+\jet$ mode have been briefly outlined in ref.~\cite{Gauld:2020ced}.
Similar ingredients are included in the computation of observables associated to the $\PWm\PH+\jet$ and $\PZ\PH+\jet$ production modes and presented here for the first time.

In the following, the relevant details for the calculation of the \DY-type and heavy quark loop induced contributions are provided in sections~\ref{D-Y} and \ref{top} respectively while
section~\ref{sigma} presents a summary of the cross section contributions for the different production modes, up to $\alphas^3$.

\subsection{Drell-Yan-type contributions}
\label{D-Y}

The Drell-Yan-type contributions are characterized by a Higgs boson coupling to an intermediate weak gauge boson and are thus proportional to $g_{VVH}^2$ at the cross-section level.
They are denoted as $\rd\sigma_{V,\;\DY}$ in the following with $V\in\lbrace\PWpm,\PZ\rbrace$ and are computed up to order $\alphas^3$ in this work.

As evident from Eq.~\eqref{eq:procs}, we consider leptonic final states from the off-shell weak gauge boson.
The amplitudes for this set of contributions can be obtained from the corresponding Drell-Yan process without the Higgs boson, provided that external momentum conservation has not been assumed in their derivation.
In that case, the off-shell \PWpm or \PZ boson can be substituted by a gauge boson that subsequently emits an on-shell Higgs particle and then decays into leptons.
The necessary Drell-Yan amplitudes for all three $V\PH+\jet$ modes up to order $\alphas^3$ can therefore be constructed from the various components of the existing calculations for the $\PW+\jet$~\cite{Gehrmann-DeRidder:2019avi} and $\PZ+\jet$~\cite{Ridder:2015dxa} processes within \nnlojet.

The above construction was directly applicable in most cases.
In a few cases, new analytic results for the amplitudes that did not impose momentum conservation on the external states were derived first.
In particular, the tree-level and two-loop amplitudes, which constitute the building blocks at the double-real and double-virtual levels to the partonic cross section were obtained in this manner.
At the real--virtual level, all one-loop amplitudes are obtained through the OpenLoops~2~\cite{Buccioni:2019sur,Buccioni:2017yxi} library (see also~\cite{Cascioli:2011va}).
This library was further used to validate the manually constructed tree-level amplitudes as well as to evaluate the four-quark tree-level amplitudes in the final computation.

As the infrared structure of the squared amplitudes is independent of the gauge-boson species, the construction of subtraction terms to treat infrared singularities is completely analogous between the various production modes of the Drell-Yan category.
Furthermore, after appropriate substitution of the colour-ordered matrix elements, all subtraction terms for the computation of the Drell-Yan-type contributions to the $V\PH+\jet$ process up to order $\alphas^3$ can be constructed from the existing $V+\jet$ calculations~\cite{Ridder:2015dxa,Gehrmann-DeRidder:2019avi} (see references therein for the appropriate antenna functions).

\subsection{Heavy quark loop induced contributions}
\label{top}

Starting from order $\alphas^2$, an additional class of corrections contribute that are characterised by a Higgs boson coupling to a heavy quark loop.
These can be treated independently from the Drell-Yan-type discussed above, as the corresponding amplitudes are separately gauge invariant.
We have included the leading-order heavy quark loop induced contributions up to $\alphas^3$ for all three production modes, using the process libraries for the relevant squared amplitudes as provided by OpenLoops~2~\cite{Buccioni:2019sur}.
These contributions are both ultraviolet and infrared finite and therefore do not require any dedicated subtraction procedure.

At order $\alphas^2$, both the charged-current and neutral-current processes receive a contribution that is characterised by the Higgs boson (but not the gauge boson) coupling to the closed quark loop---commonly referred to as \RI-type amplitudes~\cite{Brein:2011vx}.
The corresponding cross section contribution is given by the interference of the tree-level Drell--Yan-type amplitude with the \RI-type amplitude shown respectively in figures~\ref{fig:diagrams}(\subref{fig:diag_born},\subref{fig:diag_ri}).
To this end, we consider the numerically dominant top-quark loop contribution that is proportional to $\alphas^2 y_\Pqt$ and denote its contribution to the $V\PH+\jet$ cross section by $\rd\sigma_{V,\;\RI}$.
From the phenomenological point of view, taking into account only the leading-order \RI-contribution is found to be sufficient given its relatively small numerical impact, as also pointed out in ref.~\cite{Gauld:2020ced} for the case of $\PWp\PH+\jet$ production.%
 \footnote{
 	The extension of this calculation to higher order relies on accounting for all gauge-invariant sets of fermion loop contributions at a given order in $\alphas$.
 	The computation of the order $\alphas^3$ contribution relies on the knowledge of a set of heavy quark-loop amplitudes which are only partially known.
 	An illustrative example of such unknown contributions for a quark-antiquark initiated reaction is:
    	\begin{center}
       		\TopRVI{0.75cm}
   	\end{center}
}
%

In the case of the neutral-current process, additional loop-induced contributions are included that belong to the gluon-fusion channel $\Pg\Pg\to\PH\PZ\Pg$ and which are proportional to $\alphas^3 y_\Pqt^n$ with $n=0,1,2$.
These (leading order) cross section contributions are generated by squaring the gluon--gluon one-loop amplitudes, in which a gauge boson couples to either a light or a heavy-quark loop.
Sample diagrams are depicted in figure~\ref{fig:diagrams}(\subref{fig:diag_gg_top}).
In case of the left diagram, where the Higgs boson directly couples to the quark loop, we only consider the impact of the top-quark loop.
In the case of the right diagram, where the Higgs boson couples to the $\PZ$-boson and gives rise to a Drell-Yan-type amplitude, all quark flavours can circulate inside the loop.
In the following, we denote the corresponding gluon-fusion contributions to the $\PZ\PH+\jet$ production cross section as $\rd\sigma_{\PZ,\;\GGF}$.

While formally suppressed by a factor of \alphas with respect to the \RI contributions, the large gluon luminosity in LHC proton--proton collisions strongly enhances the \GGF-type corrections such that its phenomenological impact can be even more sizeable than the \RI-type terms in $\PZ\PH+\jet$ production.
They further lead to distinct features in the fiducial cross section and differential distributions (as compared to the charged-current processes, where such contributions do not exist).
These effects will be elaborated on in sections~\ref{fiducial} and \ref{distributions}.

\subsection{Cross section contributions up to order $\alphas^3$}
\label{sigma}

We conclude this section by summarizing the specific cross section contributions entering our final (N)NLO predictions.
To this end, we use a labelling based on the underlying Drell-Yan-type contributions, for which we follow the customary perturbative counting in terms of the strong coupling:
\begin{align}
   \rd\sigma^{\N{k}\LO}_{V,\;\DY}
   &=
   \sum_{i=0}^k \alphas^{i+1} \; \rd\sigma^{(i)}_{V,\;\DY}
   \, .
\end{align}

The \RI-type contributions from heavy quark loops that start to contribute from \order{\alphas^2} are consistently included (at leading order) together with \NLO corrections to the \DY-type contributions.
While the \GGF-type contributions are formally suppressed by an additional power of \alphas compared to the \RI-type contributions, we choose to include them also starting from \NLO.
The reason for this is twofold:
The large gluon luminosity at the LHC enhances this contribution to the extent that it is of similar size to both the \NLO \DY-type and \RI-type corrections.
Furthermore, the combination of the \RI- and \GGF-type contributions together with the \NLO \DY-type corrections corresponds to the state of the art prior to our work 
\cite{Luisoni:2013kna, Astill:2018ivh, Granata:2017iod, Goncalves:2015mfa}
and thus facilitates a direct comparison of the numerical impact of the newly computed \NNLO corrections to the \DY-type parts.

The labelling of (N)NLO predictions follows the definitions below:
\begin{subequations}
\label{eq:NNLO_V}
\begin{align}
   \rd\sigma^{(\N{})\NLO}_{\PW}
   &=
   \rd\sigma^{(\N{})\NLO}_{\PW,\;\DY}
   +
   \rd\sigma^{\LO(\alphas^2)}_{\PW,\;\RI}
   \, ,
   \label{eq:NNLO_W}
   \\
   \label{eq:NNLO_Z}
   \rd\sigma^{(\N{})\NLO}_{\PZ}
   &=
   \rd\sigma^{(\N{})\NLO}_{\PZ,\;\DY}
   +
   \rd\sigma^{\LO(\alphas^2)}_{\PZ,\;\RI}
   +
   \rd\sigma^{\LO(\alphas^3)}_{\PZ,\;\GGF}
   \, ,
\end{align}
\end{subequations}
and will be referred to throughout the remainder of this paper to specify the contributions included in our numerical results for fiducial cross sections and kinematic distributions.

In addition, we also provide predictions that maintain a strict power counting in \alphas when presenting the fiducial cross section results in Tables~\ref{wphfid}--\ref{zhfid}.
To this end, we define the quantity $\rd\sigma^{\NLO}_{\DY+\RI} \equiv \rd\sigma^{\NLO}_{\DY} + \rd\sigma^{\LO(\alphas^2)}_{\RI}$, which represents the \NLO cross section that only includes terms up to $\order{\alphas^2}$, i.e.\ without the \GGF-type contribution for the neutral-current process.
For consistency, we use this notation in all three tables although in the charged-current case $\rd\sigma^{\NLO}_{\PW,\;\DY+\RI} \equiv \rd\sigma^{\NLO}_{\PW}$.
In the neutral-current production mode,
this enables us to quantify the impact of the inclusion of the $\alphas^3$ corrections
arising either from the \GGF-type contributions or from the pure \DY-type contributions,
as will be further elaborated on in section~\ref{fiducial}.

\section{Numerical set-up and scale variation prescriptions}
\label{set-up}
In this section we review the calculational set-up as well as the kinematical constraints that are imposed to define the fiducial cross sections for the class of $V\PH+\jet$ processes.
We consider the associated production for all possible massive gauge bosons, $V\in\{\PWpm,\,\PZ\}$, and separately report the 1-jet inclusive ($n_\jet \ge 1$) and 1-jet exclusive ($n_\jet = 1$) cases.
Throughout the paper we will refer to these latter selections as inclusive and exclusive respectively, which strictly denote the jet multiplicity selection.

We provide predictions for proton--proton collisions at $\sqrt{s} = \SI{13}{\TeV}$ using the parton distribution function \verb|NNPDF31_nnlo_as_0118|~\cite{NNPDF:2017mvq} from the LHAPDF library~\cite{Buckley:2014ana}.
The inclusive (exclusive) processes are required to contain at least (exactly) one jet with transverse momentum $p_\bot > \SI{20}{\GeV}$.
These requirements define the \emph{inclusive} and \emph{exclusive} production cross sections, which are respectively denoted as $\sigma_{\geq1\text{j}}$ and $\sigma_{1\text{j}}$.
The anti-$k_t$ jet algorithm with the parameter $\Delta R=\num{0.5}$ is used to cluster final-state partons into jets.

Charged leptons are required to have a transverse momentum $p_{\bot,\Plpm} > \SI{25}{\GeV}$ with the rapidity constraint $|y_{\Plpm}| < \num{2.5}$.
In case of the charged-current processes, $\PWpm\PH+\jet$, an additional requirement on the minimum missing transverse energy larger than  $\SI{25}{\GeV}$ is imposed for the neutrinos.

For the electroweak input parameters, we employ the $G_\mu$-scheme with the full list of independent parameters entering the computation given by
\begin{align}
	m_{\PZ}              & = \SI{91.1876}{\GeV} ,           &
	m_{\PW}              & = \SI{80.385}{\GeV} ,            &
	\nonumber                                               \\
	\Gamma_{\PZ}         & = \SI{2.4952}{\GeV} ,            &
	\Gamma_{\PW}         & = \SI{2.085}{\GeV} ,             &
	\nonumber 						\\
	m_{\PH}			& = \SI{125.09}{\GeV} ,		&
	m_{\Pqt}^\text{pole} & = \SI{173.21}{\GeV} ,            &
	\\
	G_\text{F}           & = \SI{1.1663787e-5}{\GeV^{-2}} .
	\nonumber
\end{align}
The running of the strong coupling ($\alphas$) is evaluated using the LHAPDF library with the associated PDF set.
Finally, in the case of $\PWpm\PH$ production, a diagonal CKM matrix is used for the vector-boson--quark couplings.
We make this approximation based on the expectation that the phenomenological impact from non-diagonal effects are sub-leading for the processes under consideration.

For the central factorisation and renormalisation scale, the invariant mass of the $V\PH$ system is chosen as a characteristic hard scale of the process, $\mu_0 \equiv M_{V\PH}$.
For most kinematical distributions this choice is expected to avoid large scale-dependent logarithmic terms.

In order to estimate theoretical uncertainties from missing higher orders, we follow the conventional 7-point variation prescription:
\begin{equation*}
	\muF = \mu_0 \, \qty[ 1, \tfrac{1}{2}, 2], \qquad
	\muR = \mu_0 \, \qty[ 1, \tfrac{1}{2}, 2],
\end{equation*}
with the constraint $\tfrac{1}{2}\leq \muF/\muR \leq 2$.
All inclusive predictions adopt this error estimate for theory uncertainties.

In the case of exclusive predictions, the cross section can also be viewed as the \emph{difference} between the one- and two-jet inclusive calculation~\eqref{eq:exerr} and the above prescription corresponds to a correlated variation of the scales in the individual inclusive predictions.
In the context of exclusive results, we will therefore often refer to the standard 7-point variation as the correlated error prescription.

As a more conservative estimate, we additionally consider a second method when presenting differential distributions that
is based on ref.~\cite{Stewart:2011cf} and referred to as the uncorrelated prescription.
Here, the exclusive cross section and uncertainty measure are given as:
\begin{align}
    \label{eq:exerr}
    \sigma_{1\mathrm{j}} &\equiv \sigma_{\geq1\mathrm{j}} - \sigma_{\geq2\mathrm{j}} , &
    \Delta^2_{1\mathrm{j}} &= \Delta^2_{\geq1\mathrm{j}} + \Delta^2_{\geq2\mathrm{j}} ,
\end{align}
where $\Delta_{\geq1(2)\mathrm{j}}$ denote the uncertainties for inclusive $\PWp\PH$+1(2)-jet production obtained from their respective 7-point scale variation.
The individual uncertainties $\Delta_{\geq1\mathrm{j}}$ and $\Delta_{\geq2\mathrm{j}}$ appearing in Eq.~\eqref{eq:exerr} are each evaluated as the mean value of the upper and lower scale variation within a given kinematic selection (bin).
We choose to adopt these two error estimate prescriptions in exclusive distributions to better
assess and contrast the properties of overlapping uncertainty bands at different perturbative orders.


\section{Fiducial cross sections}
\label{fiducial}

The cross-section predictions utilizing the fiducial cuts described in section~\ref{set-up} are summarised in tables~\ref{wphfid}--\ref{zhfid} for $\PWp\PH+\jet$, $\PWm\PH+\jet$, and $\PZ\PH+\jet$ production, respectively.
In these tables, the rows correspond to results for the inclusive ($\ge$ 1 jet) (top row) and exclusive (1 jet) (bottom row) cases.
The columns of the tables correspond to fiducial cross section results at different perturbative orders including specific
contributions as detailed in section~\ref{sigma} and following the labelling introduced there.
In particular, the results considering only Drell-Yan type contributions are reported at LO in column 1, at NLO in column 2 and at NNLO in column 4. These are labelled as $\sigma^{\LO}$, $\sigma^{\NLO}_{\DY}$ and $\sigma^{\NNLO}_{\DY}$ respectively. Results for the combination of Drell-Yan type and \RI type contributions at order $\alphas^2$ are denoted by $\sigma^{\NLO}_{\DY+\RI}$ and are given in column 3.
Finally, column 5 reports the fiducial cross section results where all contributions up to order $\alphas^3$ are taken into account and are denoted by $\sigma^{\NNLO}$.
 
Only the error estimates based on the 7-point scale variation, i.e.\ the correlated prescription described in section~\ref{set-up} are quoted for the sake of clarity.
We now discuss the impact of including the NNLO \DY-type corrections for the fiducial cross section for all three associated Higgs boson production mode in the inclusive and exclusive jet cases separately.

\begin{table}
\centering
\begin{tabular}{@{\;}l S S S S S@{\;}}
\toprule
$\sigma$ $[\si{\fb}]$
& $\sigma^{\LO}$
& $\sigma^{\NLO}_{\DY}$
& $\sigma^{\NLO}_{\DY+\RI}$
& $\sigma^{\NNLO}_{\DY}$
& $\sigma^{\NNLO}$ \\
\cmidrule(r){1-1}\cmidrule{2-6}
$\PWp\PH\,+\!\ge$1 jet 
& $\num{20.99}\,^{+\num{2.09}}_{-\num{1.83}}$  
& $\num{25.80}\,^{+\num{0.87}}_{-\num{0.94}}$  
& $\num{26.12}\,^{+\num{0.94}}_{-\num{0.99}}$
& $\num{26.04}\,^{+\num{0.03}}_{-\num{0.19}}$
& $\num{26.36}\,^{+\num{0.04}}_{-\num{0.24}}$ \\
$\PWp\PH\,+$1 jet 
& $\num{20.99}\,^{+\num{2.09}}_{-\num{1.83}}$ 
& $\num{17.10}\,^{+\num{0.78}}_{-\num{1.42}}$ 
& $\num{17.42}\,^{+\num{0.73}}_{-\num{1.35}}$  
& $\num{15.27}\,^{+\num{0.54}}_{-\num{0.51}}$ 
& $\num{15.59}\,^{+\num{0.48}}_{-\num{0.44}}$ \\
\bottomrule
\end{tabular}
 \caption{The fiducial cross sections for inclusive ($\ge$ 1 jet) (top line) and exclusive (1 jet) (bottom line) associated to $\PWp\PH+\jet$ process according to the setup of section~\ref{set-up} and the cross section expressions and labelling as defined in section~\ref{sigma}. The theory error on the values represents the minimum and maximum from the 7-point scale variation. See main text for details. \label{wphfid}}
\end{table} 

\begin{table}
\centering
\begin{tabular}{@{\;}l S S S S S@{\;}}
\toprule
$\sigma$ $[\si{\fb}]$
& $\sigma^{\LO}$
& $\sigma^{\NLO}_{\DY}$
& $\sigma^{\NLO}_{\DY+\RI}$
& $\sigma^{\NNLO}_{\DY}$
& $\sigma^{\NNLO}$ \\
\cmidrule(r){1-1}\cmidrule{2-6}
$\PWm\PH\,+\!\ge$1 jet 
& $\num{12.30}\,^{+\num{1.24}}_{-\num{1.09}}$  
& $\num{15.18}\,^{+\num{0.52}}_{-\num{0.56}}$   
& $\num{15.40}\,^{+\num{0.57}}_{-\num{0.59}}$
& $\num{15.37}\,^{+\num{0.03}}_{-\num{0.12}}$    
& $\num{15.59}\,^{+\num{0.05}}_{-\num{0.15}}$ \\
$\PWm\PH\,+$1 jet 
& $\num{12.30}\,^{+\num{1.24}}_{-\num{1.09}}$  
& $\num{10.13}\,^{+\num{0.45}}_{-\num{0.83}}$
& $\num{10.35}\,^{+\num{0.41}}_{-\num{0.78}}$  
& $\num{8.61}\,^{+\num{0.44}}_{-\num{0.47}}$ 
& $\num{8.82}\,^{+\num{0.40}}_{-\num{0.43}}$ \\
\bottomrule
\end{tabular}
\caption{The fiducial cross sections for inclusive ($\ge$ 1 jet) (top line) and exclusive (1 jet) (bottom line) associated to $\PWm\PH+\jet$ process according to the setup of section~\ref{set-up} and the cross section expressions and labelling as defined in section~\ref{sigma}. The theory error on the values represents the minimum and maximum from the 7-point scale variation. See main text for details.\label{wmhfid}}
\end{table} 

\begin{table}
\centering
\begin{tabular}{@{\;}l S S S S S@{\;}}
\toprule
$\sigma$ $[\si{\fb}]$
& $\sigma^{\LO}$
& $\sigma^{\NLO}_{\DY}$
& $\sigma^{\NLO}_{\DY+\RI}$
& $\sigma^{\NNLO}_{\DY}$
& $\sigma^{\NNLO}$ \\
\cmidrule(r){1-1}\cmidrule{2-6}
$\PZ\PH\,+\!\ge$1 jet 
& $\num{5.58}\,^{+\num{0.55}}_{-\num{0.48}}$ 
& $\num{6.81}\,^{+\num{0.22}}_{-\num{0.24}}$  
& $\num{6.89}\,^{+\num{0.24}}_{-\num{0.25}}$ 
& $\num{6.92}\,^{+\num{0.02}}_{-\num{0.04}}$  
& $\num{8.05}\,^{+\num{0.42}}_{-\num{0.32}}$ \\
$\PZ\PH\,+$1 jet
& $\num{5.58}\,^{+\num{0.55}}_{-\num{0.48}}$ 
& $\num{4.49}\,^{+\num{0.22}}_{-\num{0.39}}$
& $\num{4.57}\,^{+\num{0.21}}_{-\num{0.37}}$  
& $\num{3.96}\,^{+\num{0.20}}_{-\num{0.19}}$ 
& $\num{5.08}\,^{+\num{0.27}}_{-\num{0.10}}$ \\
\bottomrule
\end{tabular}
\caption{The fiducial cross sections for inclusive ($\ge$ 1 jet) (top line) and exclusive (1 jet) (bottom line) associated to $\PZ\PH+\jet$ process according to the setup of section~\ref{set-up} and the cross section expressions and labelings at (N)NLO levels as defined in section~\ref{sigma}.
  The theory error on the values represents the minimum and maximum from the 7-point scale variation. See main text for details. \label{zhfid}}
\end{table} 

\subsection{Inclusive predictions}
We first focus our attention on the \emph{inclusive} results presented in the top row of tables~\ref{wphfid}--\ref{zhfid}.

As expected, the fiducial cross section values which involve purely Drell-Yan like contributions, display well converging perturbative behaviour as higher and higher orders are considered.
Indeed, by considering the numbers displayed in columns 2 and 4, we observe a notable stabilization of the inclusive cross section at NNLO: the correction to the central values is of $\cO(+1\%)$  while the theoretical uncertainty reduces from $\cO(4\%)$ at NLO to $\cO(1\%)$ at NNLO for all $V\PH+\jet$ processes.
Furthermore, the NNLO values are all contained within the scale uncertainty estimate of the preceding order.

The impact of the heavy quark loop contributions on the inclusive cross section can be studied by contrasting the results presented in the second and third columns (at NLO) and the fourth and fifth columns (at NNLO).
Here, the results for charged and neutral channels start to differ significantly and we observe the following behaviour:
\begin{description}
    \item[$\PW\PH+\jet$:]
        The numerical impact of the top-loop amplitudes appearing at
        order $\alphas^2$
(NLO) is relatively small and of comparable magnitude to the Drell--Yan-like $\alphas^3$ (NNLO) contribution.
This behaviour has already been noticed for the $\PWp\PH+\jet$ production mode studied in \cite{Gauld:2020ced} and highlights that the inclusion of the heavy quark loop contribution is mandatory for precision phenomenology.

From this comparison, we can further infer that the $\alphas^2$ top-loop contributions are likely to be as robust to scale variations as the NNLO Drell--Yan-like predictions and that further higher-order corrections to this contribution are only relevant in conjunction with the Drell--Yan-type corrections at \N3\LO.
The latter are however unknown and presently beyond the reach of perturbative computations.

    \item[$\PZ\PH+\jet$:]
The heavy quark loop induced contributions in $\PZ\PH+\jet$ production enter not only at order $\alphas^2$ through the \RI-type corrections, similarly to the charged-current case, but also at order $\alphas^3$ through the gluon--gluon-initiated \GGF-contributions.
The impact of heavy quark loop corrections is thus noticeably different for this Higgs boson production mode at the different orders in $\alphas$.
At $\order{\alphas^2}$, the \RI-type terms contribute to a slight increase of the cross section but have no sizeable effect on the uncertainties.
The impact of the gluon--gluon-induced $\alphas^3$ terms is positive, large in magnitude and contributes to the cross section with a residual uncertainty that is enlarged compared to the case when only Drell-Yan like contributions are included.

As noted in section~\ref{details}, the impact of the gluon--gluon-induced contributions is significantly enhanced due to the high gluon luminosity of \SI{13}{\TeV} proton--proton collisions.
These contributions form a gauge-invariant subset of the $\alphas^3$ corrections and their inclusion is justified and in fact \emph{necessary} for the correct phenomenology of the process.
Numerically, they provide the bulk of the order $\alphas^3$ corrections (i.e. they dominate over the Drell-Yan type corrections at this order), which overall leads to an $\cO(17\%)$ increase compared to the previous order and a much increased theoretical uncertainty of $\cO(5\%)$.
\end{description}

\subsection{Exclusive predictions}
Next, we consider the \emph{exclusive} fiducial cross sections given in the bottom row of the tables~\ref{wphfid}--\ref{zhfid}.

The \DY-type higher order corrections systematically suppress the fiducial cross sections as seen by comparing the numbers in the second and fourth columns of the three tables.
The size of the NNLO corrections $\cO(-10\%)$ and the associated theoretical uncertainties $\cO(5\%)$ are larger than in the inclusive case for all three Higgs boson production modes. 

The heavy quark contributions to the charged-current processes remain small at the level of $\cO(2\%)$. However, because of the \GGF-type contribution, the neutral current production process is most affected by heavy quark loop contributions. Comparing the exclusive cross sections in the last two columns of table~\ref{zhfid}, we see that the large ($\cO(13\%)$) and negative NNLO \DY-type corrections are more than compensated by the large ($\cO(25\%)$) and positive \GGF-type corrections.

We reiterate that the results shown in tables~\ref{wphfid}--\ref{zhfid} show uncertainties obtained with the correlated scale prescription. This is noted here because there is a partial cancellation in uncertainties for the exclusive $\PZ\PH+\jet$ case at the NNLO level, i.e. the last entry in the second row of table~\ref{zhfid}. If instead the correlated scale prescription is used, the uncertainty for $\PZ\PH+\jet$ at NNLO is $\Delta^{\rm NNLO}_{\rm uncorrelated} = 0.42~\fb$.

The impact of including the heavy quark loop induced contributions at the differential level, in terms of size/shape of the corrections and residual scale uncertainty at NNLO (including scale uncertainty prescriptions) for all three Higgs boson production modes, will be investigated for both inclusive and exclusive selections in section~\ref{distributions}.


\section{Differential predictions}
\label{distributions} 

In this section we turn our attention to differential observables for the three $V\PH+\jet$ processes as described in section~\ref{details} and using the numerical set-up given in section~\ref{set-up} for the \SI{13}{\TeV} LHC. We focus in particular on observables 
that are relevant for the determination of the $V\PH$ signal in experimental analyses~\cite{Aad:2020jym}.
Accordingly, we choose to separate the presentation of our results in three parts.
First, in section~\ref{sec:inc_excl}, we compare the inclusive and exclusive Higgs boson production modes for the transverse momentum spectra of Higgs boson and vector boson for the $\PWm\PH+\jet$, and $\PZ\PH+\jet$ processes.  Second, and motivated by the necessity to suppress overwhelming multi-jet backgrounds from $\Pqt\Paqt$ production where one of the top quarks decays semi-leptonically, we present results for the {\em exclusive} charged current production modes in section~\ref{sec:CC_excl}.  Third, we discuss results for the {\em inclusive} neutral current production mode in section~\ref{sec:NC_inc} because typically any number of jets is accepted in the experimental analysis to increase the signal acceptance.

For all results presented in section~\ref{distributions}, the computation of the differential cross sections is performed using the expressions presented in section~\ref{sigma}.
More specifically, the (N)NLO expressions are given in 
eq.~\eqref{eq:NNLO_W} for the charged current case and in eq.~\eqref{eq:NNLO_Z} for the neutral current process.  In other words, the previously known heavy quark \RI-type contributions starting at order $\alphas^2$ and, in case of the neutral-current process, the gluon fusion contribution appearing first at $\cO(\alphas^3)$ are included in both NLO and NNLO predictions.
With this choice, the study of the impact of the pure NNLO (i.e $\cO(\alphas^3)$) \DY-type contribution computed for the first time here is facilitated. 

\subsection{Inclusive vs.\ exclusive predictions}
\label{sec:inc_excl} 
Our aim here is to compare the results obtained for predictions in the 1-jet inclusive ($n_\jet \ge 1$) and 1-jet exclusive ($n_\jet = 1$) categories at a given fixed order in $\alphas$.
We shall here focus on the transverse momentum spectra of the Higgs boson and the vector boson 
in the production processes $\PWm\PH+\jet$, and $\PZ\PH+\jet$ and highlight similarities and differences for these two
production modes. 
We begin our discussion with a general description of the figures illustrating these differential results. 
Specifically, in section~\ref{sec:inc_excl}, the figures illustrating the results are composed of four sub-panels:
The top panel shows the absolute predictions; the two centre panels show the $K$-factors for the inclusive and exclusive processes, respectively.
For the exclusive case, we include the two error estimates as introduced in section~\ref{set-up} that constitute the correlated (shaded fill) and the uncorrelated (solid fill) prescription.
Lastly, the bottom (fourth) panel displays the veto efficiency defined as
\begin{align*}
    \epsilon_\text{veto}(\mathcal{O}) =
    \frac{\mathrm{d}\sigma_{1\text{j}}     / \mathrm{d}\mathcal{O}}
         {\mathrm{d}\sigma_{\geq1\text{j}} / \mathrm{d}\mathcal{O}}
         ,
\end{align*}
for an observable $\mathcal{O}$ and for which the uncorrelated error estimate of the exclusive 1-jet cross section is used.
For the uncorrelated scale prescription as defined in Eq.~\eqref{eq:exerr}, the uncertainties from the numerator and denominator are combined by applying error propagation.

\begin{figure}
   \centering
   \begin{subfigure}[h]{.4\textwidth}
      \includegraphics[width=\textwidth]{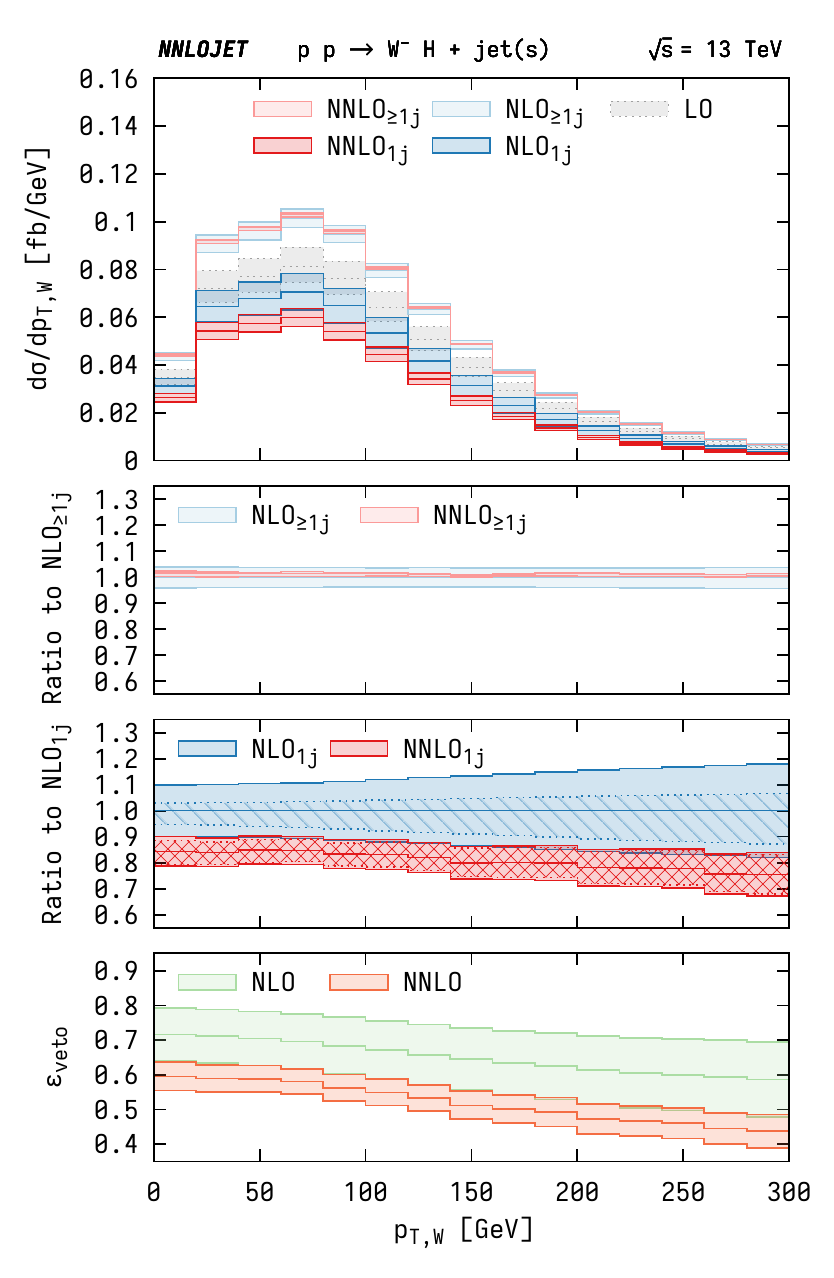}
      \subcaption{}
      \label{fig:4panel_wmptw}
   \end{subfigure}
   \begin{subfigure}[h]{.4\textwidth}
      \includegraphics[width=\textwidth]{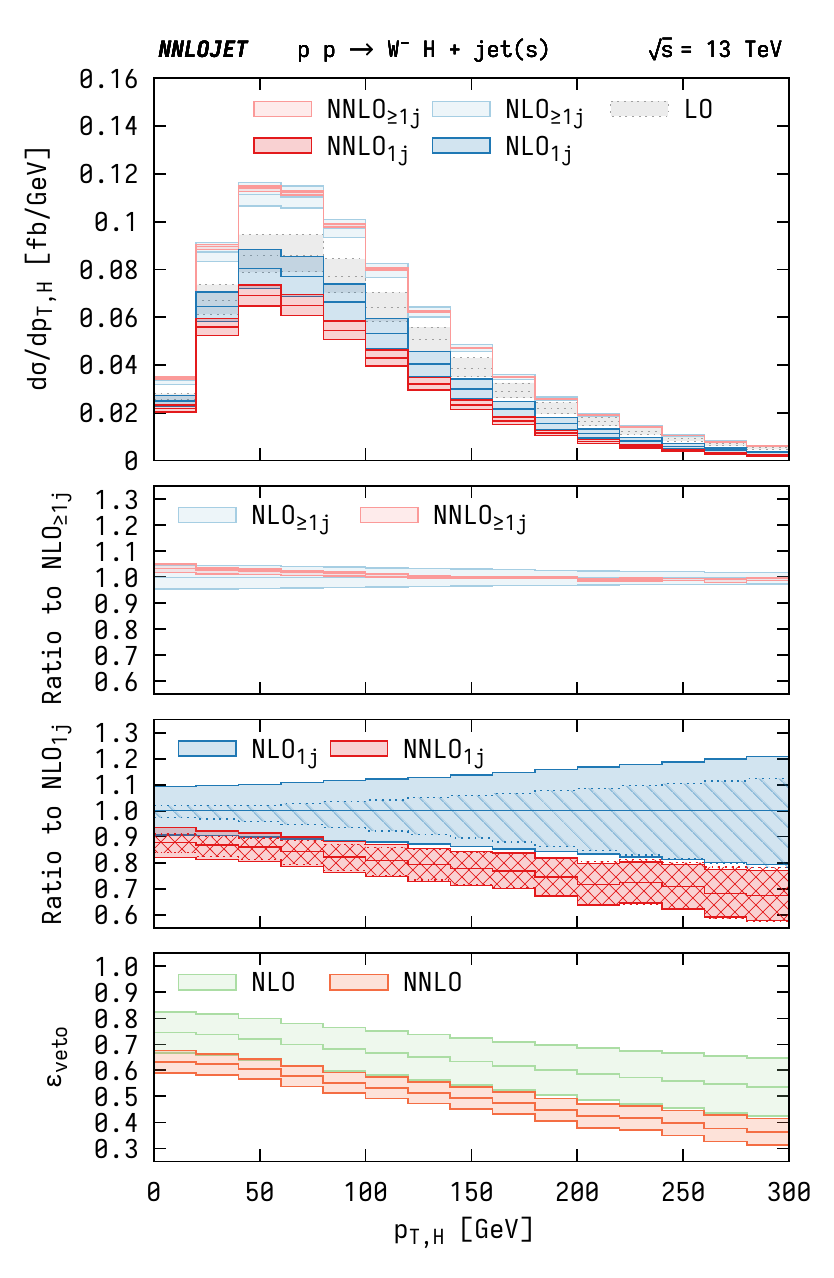}
      \subcaption{}
      \label{fig:4panel_wmpth}
   \end{subfigure}
\caption{Transverse momentum distribution of (a) the gauge boson and (b) the Higgs boson for the $\PWm\PH+\jet$ process.} 
   \label{fig:4panel_wm}
\end{figure}

\begin{figure}
   \centering
   \begin{subfigure}[h]{.4\textwidth}
      \includegraphics[width=\textwidth]{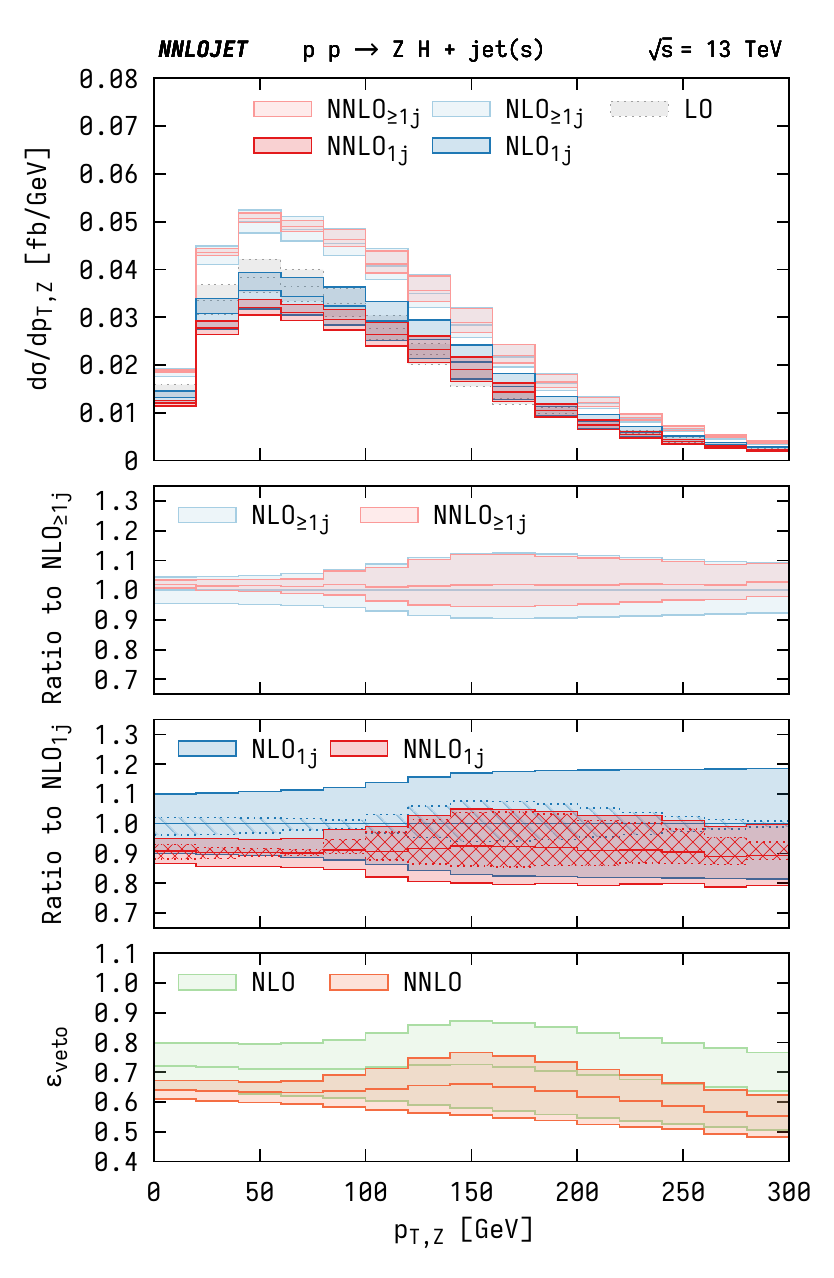}
      \subcaption{}
      \label{fig:4panel_zptw}
   \end{subfigure}
   \begin{subfigure}[h]{.4\textwidth}
      \includegraphics[width=\textwidth]{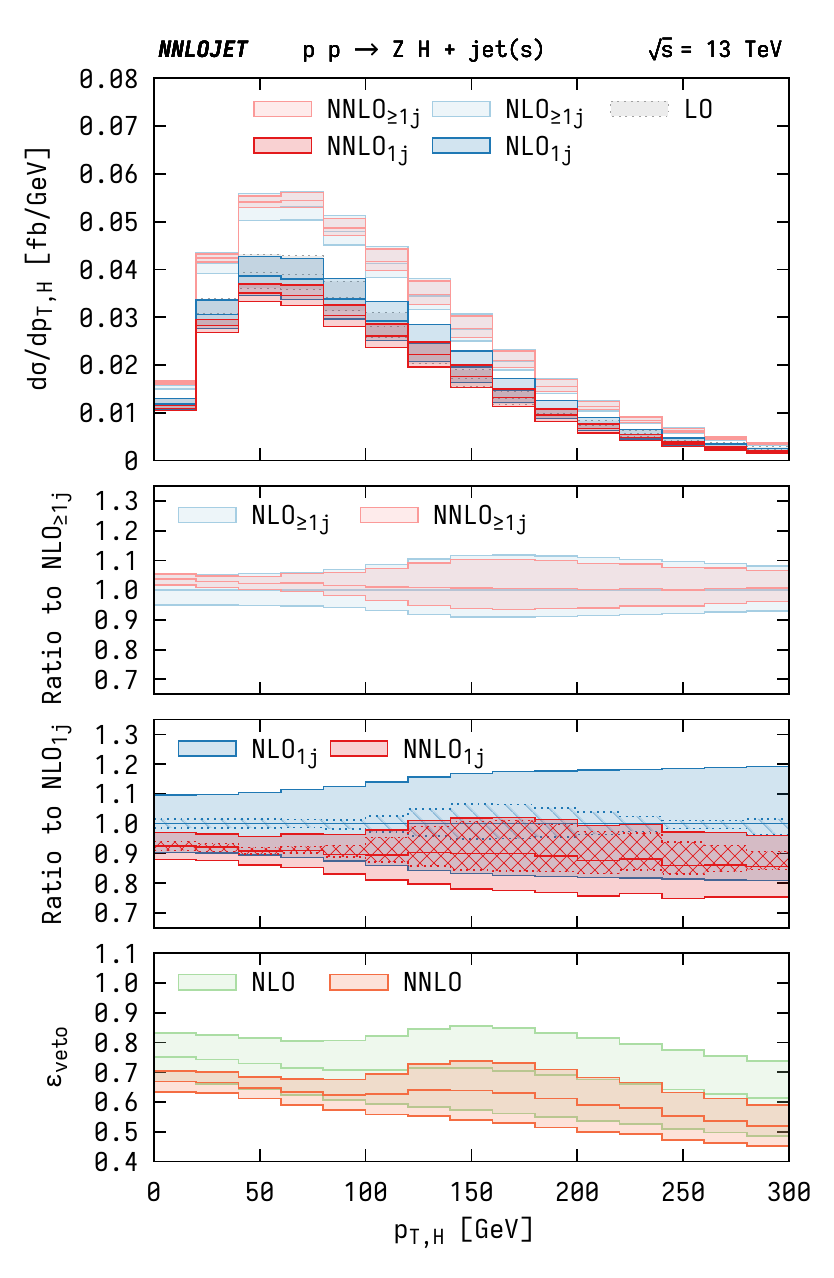}
      \subcaption{}
      \label{fig:4panel_zpth}
   \end{subfigure}
\caption{Transverse momentum distribution of (a) the gauge boson and (b) the Higgs boson for the $\PZ\PH+\jet$ process.} 
   \label{fig:4panel_z}
\end{figure}

Figure~\ref{fig:4panel_wm} and  figure~\ref{fig:4panel_z} present 
predictions for the transverse momentum distribution of the gauge boson and Higgs boson for the $\PWm\PH+\jet$ and the $\PZ\PH+\jet$ processes including corrections up to order ${\cal O} (\alphas^3)$ in the computations. 

We observe that, for both charged- and neutral-current cases shown in figures~\ref{fig:4panel_wmptw} and \ref{fig:4panel_zptw}
the gauge boson distributions exhibit a maximum slightly above $p_\rT\sim50~\GeV$, which is induced by the minimum transverse momentum cut on the lepton (and $E_{\bot,\text{miss}}$ for the charged-current case) of $25~\GeV$.
The requirement of an additional resolved jet relaxes the otherwise strict constraint $p_\rT^V > 50~\GeV$ for a back-to-back $V\PH$ configuration at Born level without a jet.
Nonetheless, the dominant kinematic configuration is still associated with an approximate back-to-back configuration between \PH and $V$ (see also Fig.~\ref{fig:z_deltaphi}), as suggested by the peak location around $50~\GeV$, displayed in the first panels of these figures.

An observation that further supports this conclusion is the qualitative similarity of the Higgs boson transverse momentum distribution shown in figures~\ref{fig:4panel_wmpth} and \ref{fig:4panel_zpth} for neutral and charged boson production, respectively
as compared to the transverse momentum spectrum of the gauge boson, depicted in figures~\ref{fig:4panel_wmptw} and \ref{fig:4panel_zptw}. 
While no selection cuts are directly enforced on the Higgs boson, restrictions imposed by fiducial cuts on the gauge boson indirectly translate to the Higgs boson, in particular, with a maximum that aligns well with that seen in the respective gauge boson distribution.
As a result, higher-order corrections to the two observables display quantitatively similar features for both charged and neutral production modes, as best seen comparing qualitatively the first panels in figure~\ref{fig:4panel_wm} and    figure~\ref{fig:4panel_z}. It is seen that, for all distributions above the peak value of  $50~\GeV$, the cross section decreases monotonically towards higher values of the transverse momentum of the boson produced. 

As a general observation, the inclusive predictions receive positive higher order corrections while corrections to the exclusive predictions are negative. This is evident in the first panel of figures~\ref{fig:4panel_wmptw} and \ref{fig:4panel_zptw} where the NNLO inclusive distributions (light red) are larger than the corresponding NLO inclusive distributions (light blue), while the reverse is true for the exclusive distributions.
Further, we note that the inclusion of higher-order corrections has a bigger impact in the exclusive rather than in the inclusive case, which was noticed for the fiducial cross sections presented in section~\ref{fiducial}.
It can be seen clearly for the distributions when comparing the size and uncertainties of the NNLO $K$-factors shown in the second and third panels, as well as the veto efficiency results presented in the bottom panel.
For the $\epsilon_{\rm veto}$ distributions shown in figure~\ref{fig:4panel_wmptw}, the overlap of the theory uncertainties at NLO and NNLO is marginal, as a result of these large corrections to the exclusive case.

\subsubsection{$\PWm\PH+\jet$ distributions}
Focussing on the distributions for the charged-current process shown in figure~\ref{fig:4panel_wm},  we observe that the NNLO inclusive $K$-factors (shown in the second panels) for both spectra are largely flat and close to unity with percent level residual scale uncertainties. 

Instead, in the exclusive case (as shown in the third and fourth panels) the size of the corrections are considerably larger.  The ratio of NNLO to NLO (red bands) decreases from $\cO(-15\%)$ at low transverse momentum to $\cO(-25\%)$ at larger transverse momentum values, while the residual scale uncertainty band is roughly decreased by a factor of two.
%
Furthermore the NNLO and NLO uncertainty bands only overlap if the uncorrelated scale variation prescription is used. The veto efficiency, shown in the fourth panel also decreases towards higher values of $p_\rT$, indicating  that the high transverse momentum region is dominated by VH +2-jet production, which is present at NLO level only in these computations. 

In both exclusive and inclusive cases, the inclusion of the $\cO(\alphas^3)$ \DY-type corrections, stabilises the predictions for the transverse momenta spectra, and leads to a reduction of the residual theoretical uncertainties. This is most pronounced in the exclusive production mode, where the inclusion of the NNLO corrections has a bigger impact.

\subsubsection{$\PZ\PH+\jet$ distributions}

For the distributions for the neutral production mode presented in figure~\ref{fig:4panel_z}, we observe that the impact of the inclusion of the NNLO corrections has a similar qualitative effect on the inclusive and exclusive $K$-factors. 
As in the charged-current case, we observe that the size of these $K$-factors in the exclusive category is larger than in the inclusive category. This is clearly seen in the second and third panels, together with the veto efficiency in the fourth panel.  

In addition, we notice a distortion of the shape of the $K$-factors for both distributions which is absent in the charged current mode. This is present in both inclusive and exclusive jet categories:  Up to $p_\rT \gtrsim 140~\GeV$ the \DY-type NNLO corrections are at the percent level in the inclusive case and ten to twenty percent in the exclusive case. The NLO and NNLO uncertainty bands also nicely overlap. Above this threshold we observe a clear change in size and shape of the $K$-factors and efficiencies. 
The shape distortion and resulting inflation of the theoretical uncertainty band above $p_\rT \gtrsim 140~\GeV$ corresponds to a change in the dominant contributions participating at $\cO(\alphas^3)$. 
Indeed, this onset of large positive corrections corresponds to the $\Pqt\Paqt$ threshold in loop diagrams such as the one shown in figure~\ref{fig:diag_gg_top} and marks the transition from the low transverse momentum region where \DY-type contributions are dominant and where the cross section is the largest, to the high-$p_\rT$ regime where quark loop contributions induced by the \GGF-type of processes also have a sizeable impact.
The \GGF-type corrections are kinematically in a leading-order configuration, and therefore influence the inclusive and exclusive predictions in a similar manner. The veto efficiency, shown in the fourth panel of these figures exhibits the same qualitative behaviour as the $K$-factors for both distributions.
 
The impact of the \DY- and non-\DY-type corrections present at $\cO(\alphas^3)$ will be further analysed in section~\ref{sec:NC_inc} for a wider set of observables in the inclusive neutral production mode.

\subsection{The exclusive charged-current processes}
\label{sec:CC_excl}
In this section, we focus on the 1-jet exclusive $\PW H$ process.
As previously noted, the motivation for considering a 1-jet exclusive selection is that the large background from the semi-leptonic $\Paqt\Pqt$ channel can be suppressed to a large degree.
Our aim is to compare and contrast the effect of the NNLO corrections on the differential predictions for the two charged-current processes.
We consider four observables: the transverse momenta spectra of the vector boson and the Higgs boson, as well as the transverse momentum and rapidity distribution of the jet shown in figures~\ref{fig:cc_ptw}--\ref{fig:cc_yj1}.
These figures contain two panels: 
The top panel shows the absolute distributions at each perturbative order in $\alphas$ and the lower panel displays the respective $K$-factors. In the latter, the denominator of the NLO ratio contains \DY type and heavy quark loop $R_I$ type contributions up to order $\alphas^2$ as detailed in section~\ref{sigma}.

\subsubsection{The transverse momentum spectra of vector and Higgs bosons} 

 \begin{figure}
   \centering
   \begin{subfigure}[h]{.4\textwidth}
      \includegraphics[width=\textwidth]{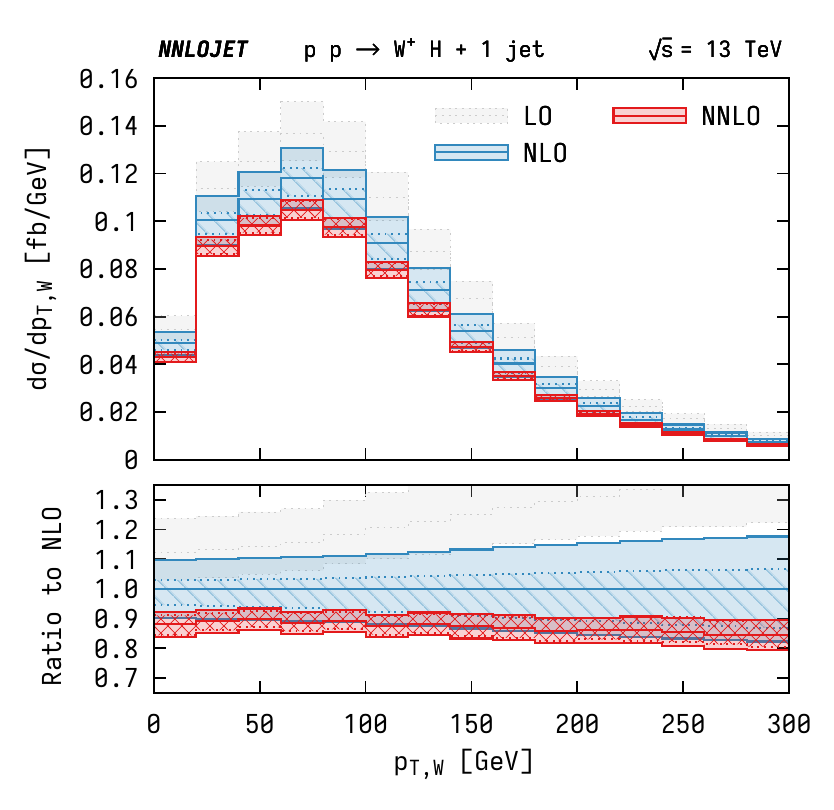}
      \subcaption{}
      \label{fig:wp_ptw}
   \end{subfigure}
   \begin{subfigure}[h]{.4\textwidth}
      \includegraphics[width=\textwidth]{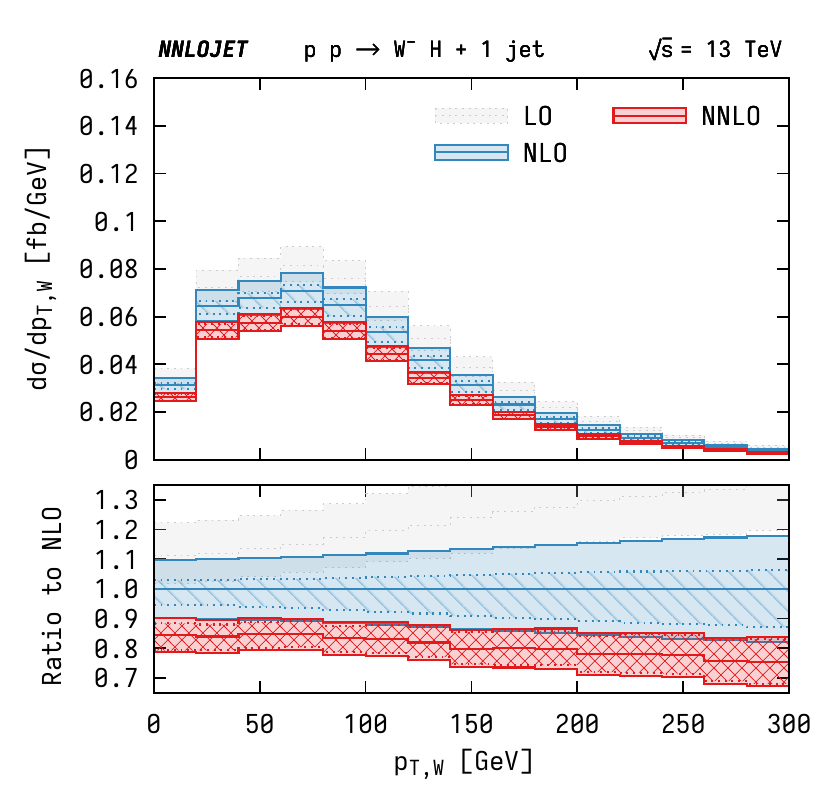}
      \subcaption{}
      \label{fig:wm_ptw}
   \end{subfigure}
   \caption{The transverse momentum distribution of the W boson produced in exclusive $\PW\PH+\jet$ production for (a) \PWp (b)  \PWm.}
   \label{fig:cc_ptw}
\end{figure}

 \begin{figure}
   \centering
   \begin{subfigure}[h]{.4\textwidth}
      \includegraphics[width=\textwidth]{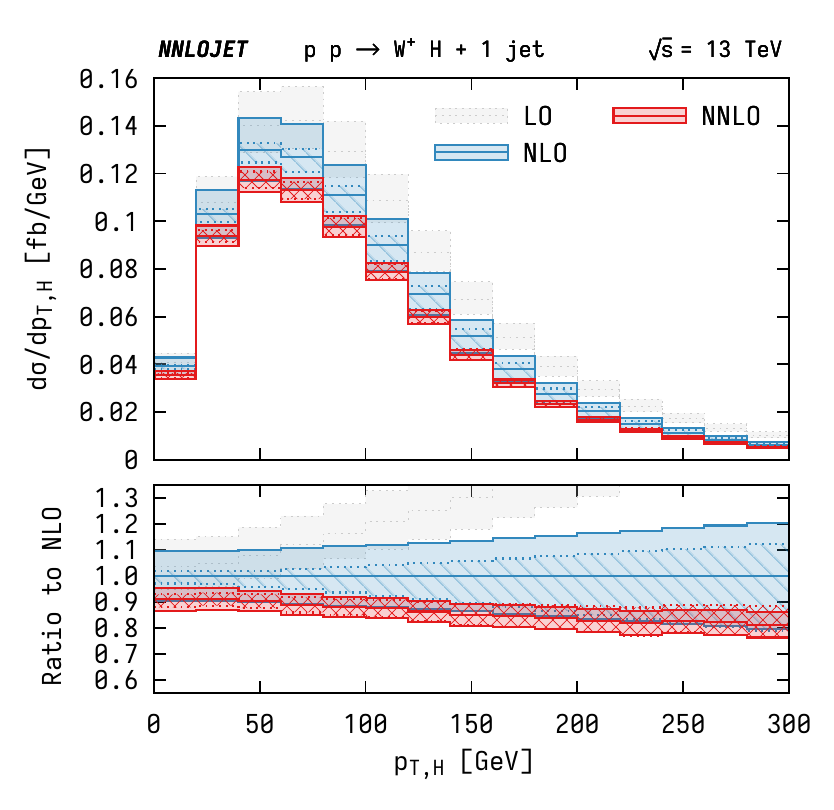}
      \subcaption{}
      \label{fig:wp_pth}
   \end{subfigure}
   \begin{subfigure}[h]{.4\textwidth}
      \includegraphics[width=\textwidth]{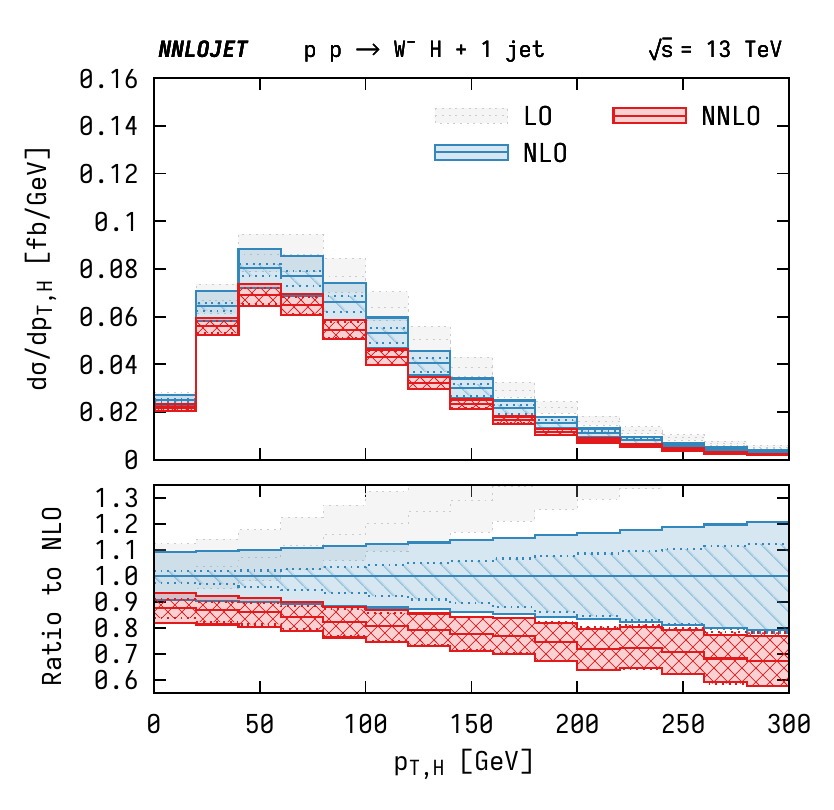}
      \subcaption{}
      \label{fig:wm_pth}
   \end{subfigure}
   \caption{The transverse momentum distribution of the Higgs boson produced in exclusive $\PW\PH+\jet$ production for (a)  \PWp  (b) \PWm.}
   \label{fig:cc_pth}
\end{figure}

The behaviour of both the gauge boson and the Higgs boson transverse momentum spectra for the $\PWm\PH+\jet$ process were analysed in section~\ref{sec:inc_excl}. A qualitatively similar behaviour is observed for the case of the production of a positively
charged boson as shown in left part (first panel) of figures~\ref{fig:cc_ptw} and \ref{fig:cc_pth} and was already noticed in \cite{Gauld:2020ced}.

To compare the two production modes, we therefore focus on the $K$-factors presented in the second panels of figures~\ref{fig:cc_ptw} and \ref{fig:cc_pth}.
We see that the corrections are sizeable and only the more conservative uncorrelated error estimate is able to provide overlapping uncertainty bands when going from NLO to NNLO.
We further observe that the NNLO corrections and the associated uncertainties at NNLO are always larger in the $\PWm$ case (up to 20\%) compared to the $\PWp$ case (which are about 10\%).
A possible interpretation is that the relative contribution of the $\Pg\Pg$ and $\Pq\Pq'$ partonic channels is larger for the $\PWm$ mode, and this results in larger theoretical uncertainties. This is particularly true for the regions of kinematic distributions which are sensitive to the input PDFs at large-$x$ values (i.e.\ forward rapidities), where the $\Pqd/\Pqu$ ratio becomes small and the impact of these NLO channels increases.

\subsubsection {Jet related observables}
\begin{figure}
   \centering
   \begin{subfigure}[h]{.4\textwidth}
      \includegraphics[width=\textwidth]{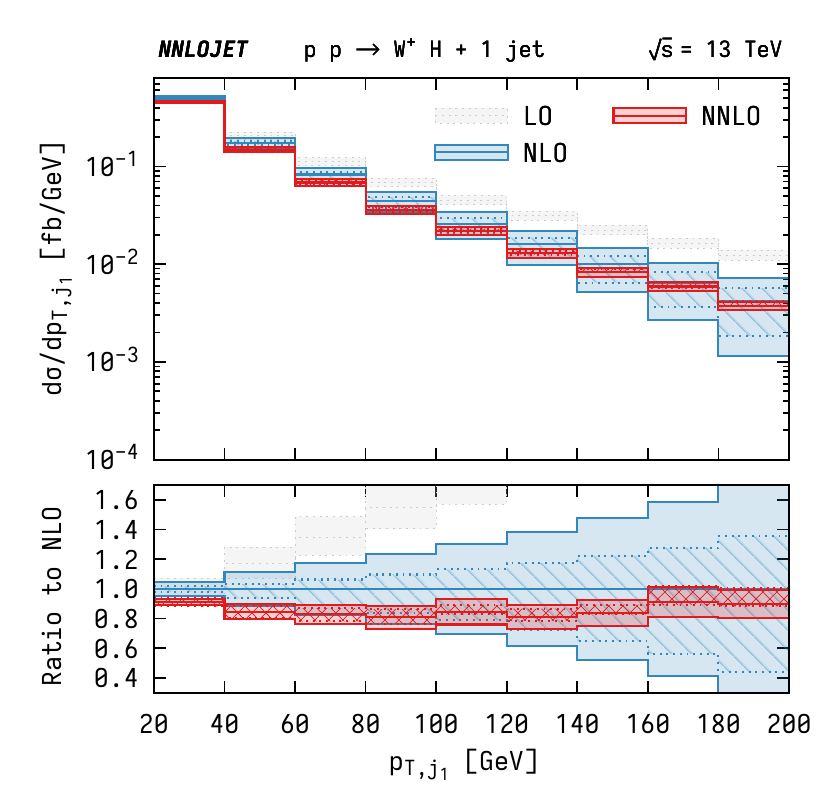}
      \subcaption{}
      \label{fig:wp_ptj1}
   \end{subfigure}
   \begin{subfigure}[h]{.4\textwidth}
      \includegraphics[width=\textwidth]{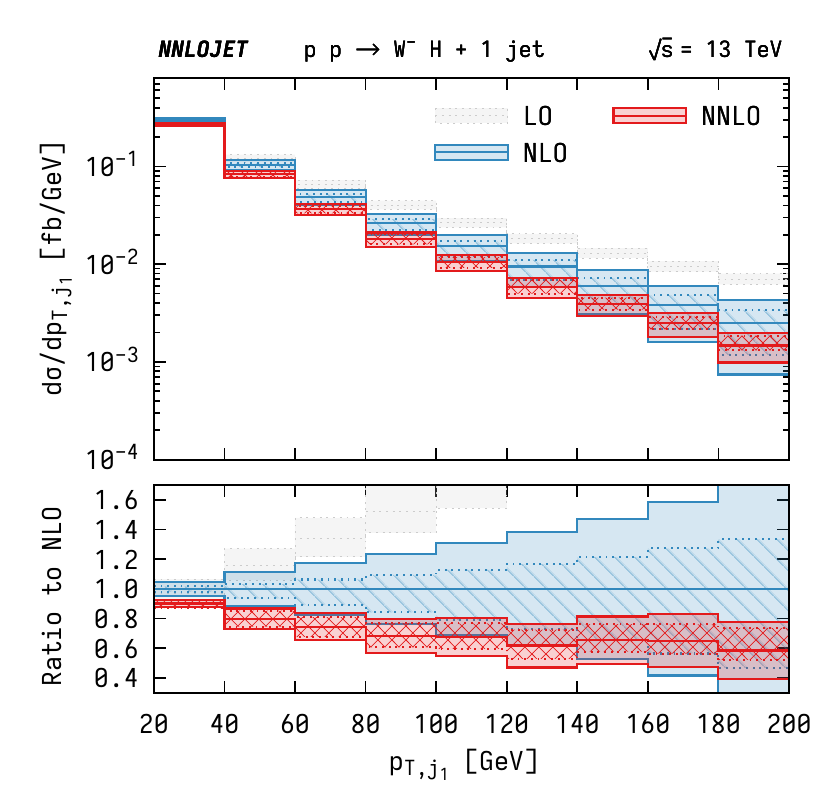}
      \subcaption{}
      \label{fig:wm_ptj1}
   \end{subfigure}
   \caption{The transverse momentum distribution of the jet produced in exclusive $\PW\PH+\jet$ production for (a)  \PWp  (b)  \PWm.}
   \label{fig:cc_ptj1}
\end{figure}

\begin{figure}
   \centering
   \begin{subfigure}[h]{.4\textwidth}
      \includegraphics[width=\textwidth]{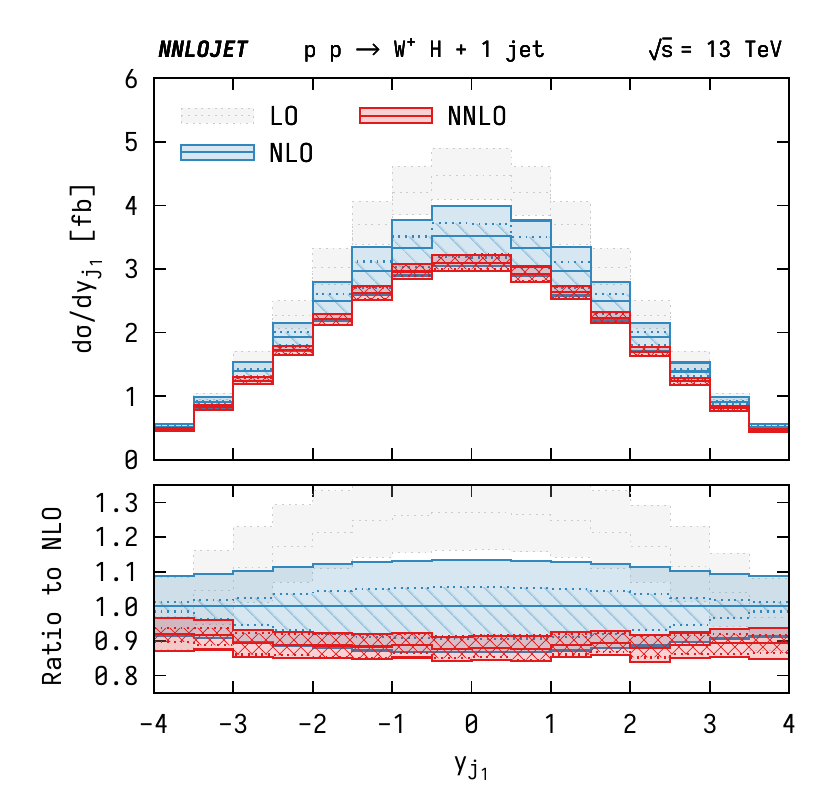}
      \subcaption{}
      \label{fig:wp_yj1}
   \end{subfigure}
   \begin{subfigure}[h]{.4\textwidth}
      \includegraphics[width=\textwidth]{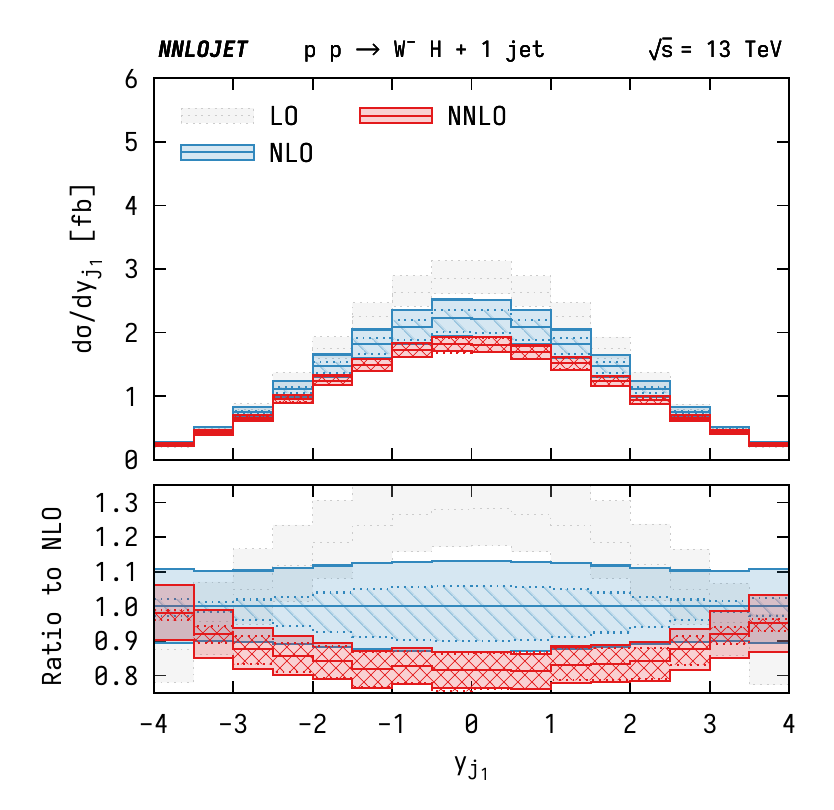}
      \subcaption{}
      \label{fig:wm_yj1}
   \end{subfigure}
   \caption{The rapidity distribution of the jet produced in exclusive $\PW\PH+\jet$ production for (a) \PWp  (b) \PWm.}
   \label{fig:cc_yj1}
\end{figure}

We now consider predictions for two specific jet related exclusive observables, 
namely, the transverse momentum and the rapidity of the leading jet presented in figures~\ref{fig:cc_ptj1} and \ref{fig:cc_yj1} respectively.

For both of these exclusive Higgs boson production modes, the inclusion of the NNLO corrections lead to a stabilisation of the predictions with reduced left-over theoretical uncertainty bands compared to the results obtained at NLO. However, the NNLO corrections to both the transverse momentum distribution and the rapidity distribution of the leading jet are qualitatively different for the two charged current processes.  

As shown in figure~\ref{fig:cc_ptj1}, the inclusion of the NNLO corrections lead to a stabilisation of the predictions with reduced left-over theoretical uncertainty bands compared to the results obtained at NLO for both of these exclusive Higgs boson production modes.  However, the quantitative impact is different for the \PWp and \PWm cases.   Compared to NLO, the NNLO corrections to the transverse momentum distribution for the negatively charged process lead to a reduction of $\cO(-10\%)$ at low transverse momentum that reaches $\cO(-40\%)$ for $p_{\rT,\jet}$ = 200~GeV where the production of two high $p_\rT$ jets takes over and the exclusive process with exactly one jet in the final state is consequently strongly suppressed. For \PWp production, the corrections depend more weakly on the jet transverse momentum, and are of $\cO(-10\%)$ at both low $p_{\rT,\jet}$ and high $p_{\rT,\jet}$, and reach $\cO(-20\%)$ for intermediate values of $p_{\rT,\jet}$.
In both cases, the residual scale uncertainties are considerably reduced at NNLO, and are typically $\pm10\%$ for the production of a \PWp and $\pm20\%$ for the \PWm case (in the regime of high $p_{\rT,\jet}$ values).

The results for the rapidity distribution of the leading jet shown in figure~\ref{fig:cc_yj1} also exhibit some marked differences between the \PWp and \PWm production modes.
The most noticeable differences occur in the forward rapidity region, where larger NNLO $K$-factors and residual uncertainties in the \PWm case are observed. 

In both cases, the NNLO $K$-factors relative to NLO, are negative for centrally produced jets, $-12\%$ for \PWp and $-18\%$ for \PWm.  However, the NNLO corrections are largely independent of $y_{j_1}$ for the positive charged-current process, while for the $\PWm\PH+\jet$ process, the NNLO corrections are small at large $|y_{j_1}|$. In both cases, the residual scale uncertainties are considerably reduced at NNLO, and are typically $\pm6\%$ for the production of a \PWp and $\pm8\%$ for the \PWm case.

The difference in behaviour can be attributed to the parton distribution functions and the differences in the luminosity of the partons that initiate the hard scattering.
At large rapidities, high momentum fractions $x$ are probed that can vary significantly between the two charged-current processes where valence- and sea-quark contributions are interchanged.
Note that similar qualitative features were also observed when comparing the $\PWp+\jet$ and $\PWm+\jet$ processes (with no Higgs boson in the final state) for the leading-jet rapidity distributions in ref.~\cite{Gehrmann-DeRidder:2019avi}.

\subsection{The inclusive neutral-current process}
\label{sec:NC_inc}

In this section, we present differential predictions related to the associated production of a Higgs boson, a neutral $Z$ boson 
and at least one hadronic jet in the final state, i.e 1-jet inclusive production ($n_\jet \geq 1$).
We focus here on three categories of inclusive  observables: (a) the transverse momenta of the Higgs boson and the Z boson (figure~\ref{fig:nc_zh}),
(b) the leading-jet related observables (figure~\ref{fig:nc_jet})
and (c) those specific observables participating explicitly in the $\PZ\PH$ signal extraction (figure~\ref{fig:nc_more}) in an experimental context~\cite{ATLAS:2020fcp}.
Each figure has a common structure with the top panel showing the absolute distributions at each perturbative order in $\alphas$ while the bottom panel displays the ratio to NLO. The ingredients (and labelling) of the cross section results at the 
$k$-th order in $\alphas$ are in-line with the expressions presented in section~\ref{sigma}. In particular, it is worth recalling that the gluon-gluon fusion contribution is included here in both the NLO and NNLO cross section expressions,  
as defined in eq.~\eqref{eq:NNLO_Z}.
 
A main goal of this section is to study the impact of the inclusion of the different contributions entering at $\cO(\alphas^3)$ in different kinematical regions associated with the particular inclusive observable under consideration.
To best achieve this goal, and better identify which higher order corrections have the highest phenomenological impact, we also show the results obtained using purely the \DY--type contributions at NLO level denoted by NLO$_{\DY}$. 

\subsubsection{The Z boson and Higgs boson transverse momenta distributions} 
\begin{figure}
 \centering
   \begin{subfigure}[h]{.4\textwidth}
      \includegraphics[width=\textwidth]{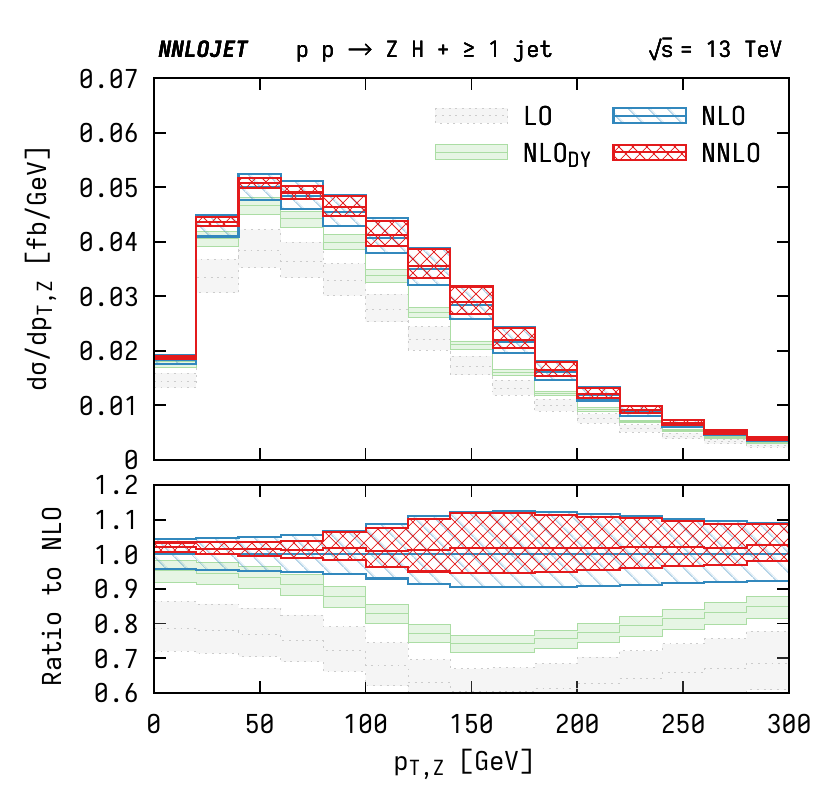}
      \subcaption{}
      \label{fig:z_ptw}
   \end{subfigure}
   \begin{subfigure}[h]{.4\textwidth}
      \includegraphics[width=\textwidth]{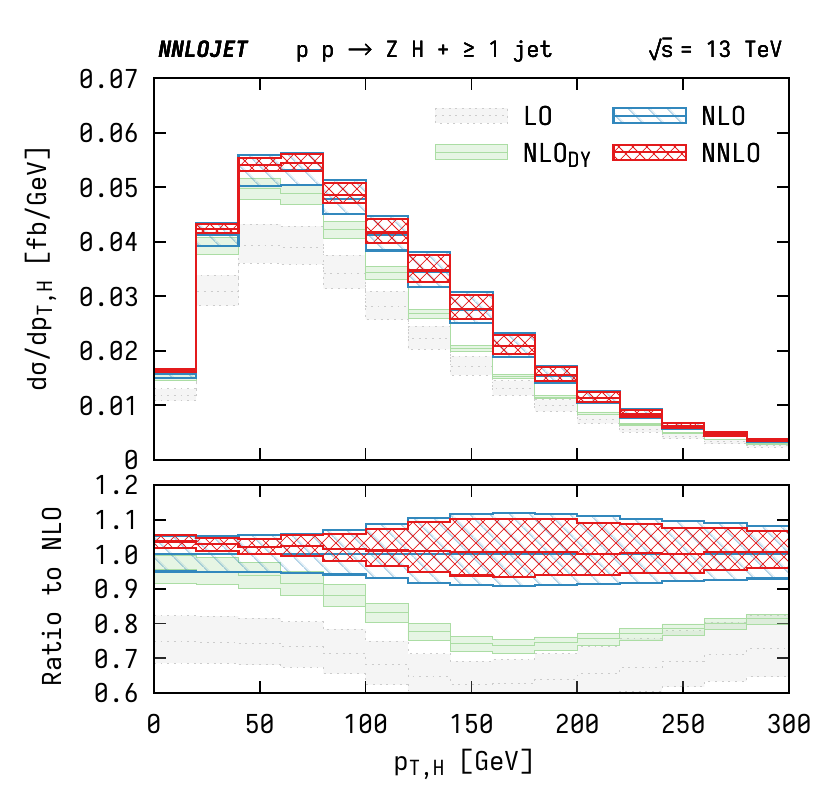}
      \subcaption{}
      \label{fig:z_pth}
   \end{subfigure}
   \caption{The transverse momentum of (a) the Z-boson and (b) the Higgs boson.}
   \label{fig:nc_zh}
\end{figure}

As discussed in section~\ref{sec:inc_excl}, the qualitative behaviour of both ($\PZ$ boson and Higgs boson) spectra presented in the first panels of figures~\ref{fig:z_ptw} and \ref{fig:z_pth} shows that the distributions have a peak value close to $p_\rT^V > 50~\GeV$ and then falls off sharply for increasing values of $p_\rT$. The $K$-factors show that up to $p_\rT \sim 140~\GeV$, the NNLO corrections are small and positive (at the percent level) and the NNLO uncertainty band lies within the NLO uncertainty. At higher transverse momenta, the size of the NNLO corrections increases and the scale uncertainties also increase.
This reflects the impact of the gluon-gluon fusion contributions, (labelled as \GGF type, see section~\ref{details}), which are becoming phenomenologically relevant for these large values of $p_\rT$ and induce sizeable changes in the transverse momenta spectra. As these contributions first appear at leading order at $\cO(\alphas^3)$, this behaviour is not unexpected.  Their impact is clearly seen by comparing the results denoted as NLO (where both \DY and \GGF type contributions included) with those labelled as NLO$_{\DY}$.

\subsubsection{Jet observables in ZH +jet production} 

\begin{figure}
   \centering
   \begin{subfigure}[h]{.4\textwidth}
      \includegraphics[width=\textwidth]{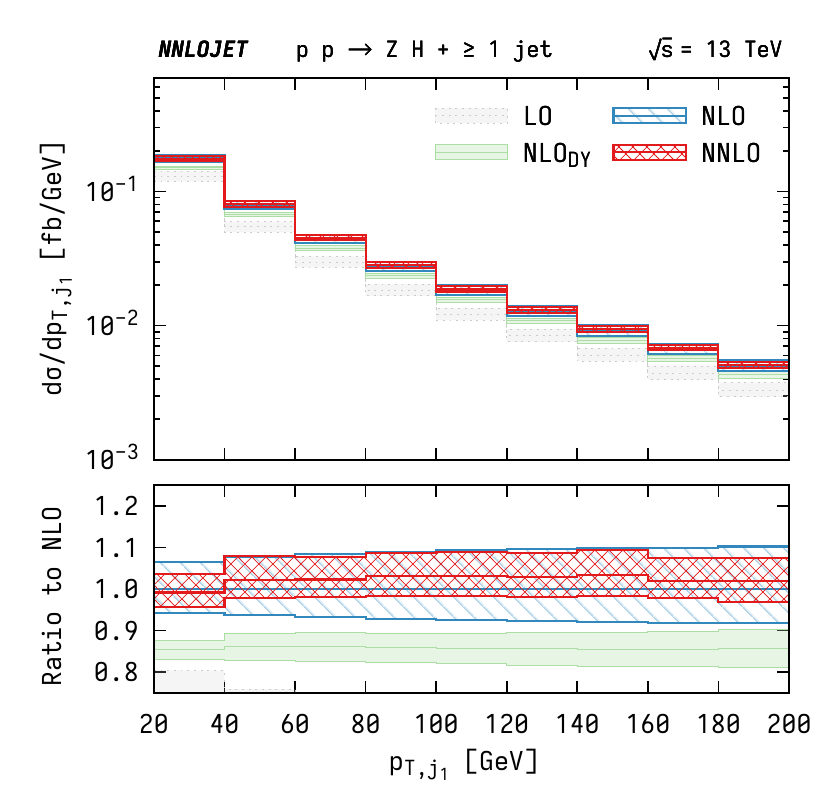}
      \subcaption{}
      \label{fig:z_ptj1}
   \end{subfigure}
   \begin{subfigure}[h]{.4\textwidth}
      \includegraphics[width=\textwidth]{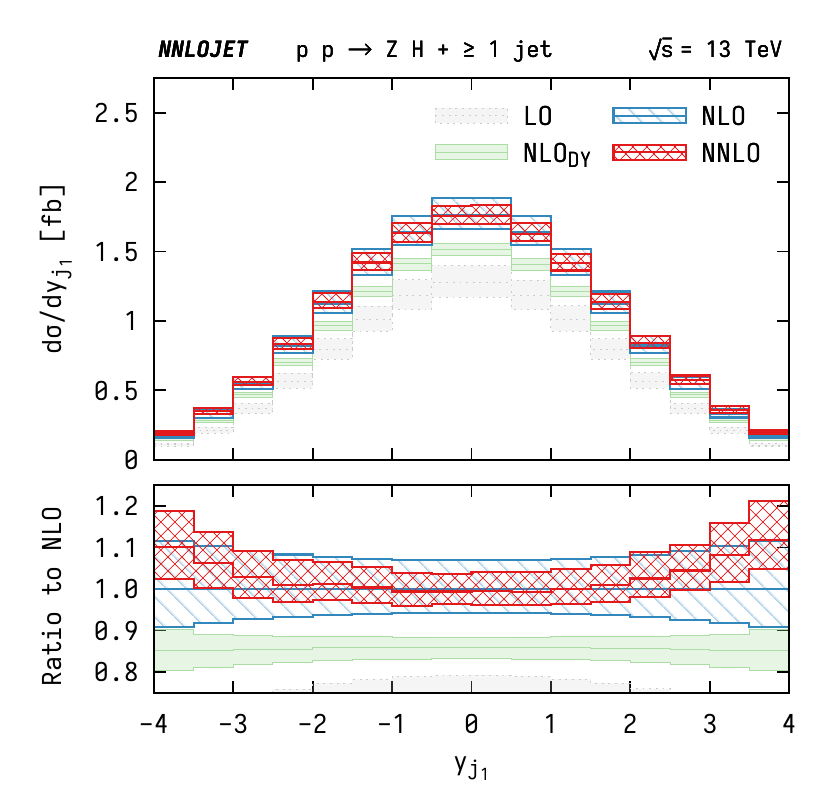}
      \subcaption{}
      \label{fig:z_yj1}
   \end{subfigure}
   \caption{Leading jet observables: (a) transverse momentum and (b) rapidity.}
   \label{fig:nc_jet}
\end{figure}

Focussing first on the leading jet $p_\rT$ spectrum displayed in figure~\ref{fig:z_ptj1}, we observe that the NNLO corrections are (a) mostly independent of $p_{\rT,j}$ and at the percent level, (b) clearly lead to a stabilization of the perturbative predictions with scale uncertainties that are contained within the NLO uncertainty band, and (c) have residual scale uncertainties of about $\pm 8$\% compared to NLO, over the whole kinematical range.
 
The leading jet rapidity distribution is shown in figure~\ref{fig:z_yj1}. We see that the size of NNLO corrections and corresponding scale uncertainty band are considerably larger than those observed for the leading jet transverse momentum observable. From the second panel, we observe that compared to NLO, the NNLO prediction is small in the central rapidity region but increases with $|y_{j_1}|$, with little reduction of scale uncertainty band. 

Comparing the NLO and NLO$_{\DY}$ results leads us to identify the inclusion of  \GGF-type loop contributions as significant (about $+20\%$ compared to NLO) and uniform in $p_{\rT,j}$ and $y_{j_1}$. The inclusion of the \GGF contribution changes the phenomenological picture at NNLO specially in the most forward rapidity region, where the cross section is the smallest.
   
\subsubsection{ZH +jet specific discussion} 

\begin{figure}
  \centering
  \begin{subfigure}[h]{.4\textwidth}
    \includegraphics[width=\textwidth]{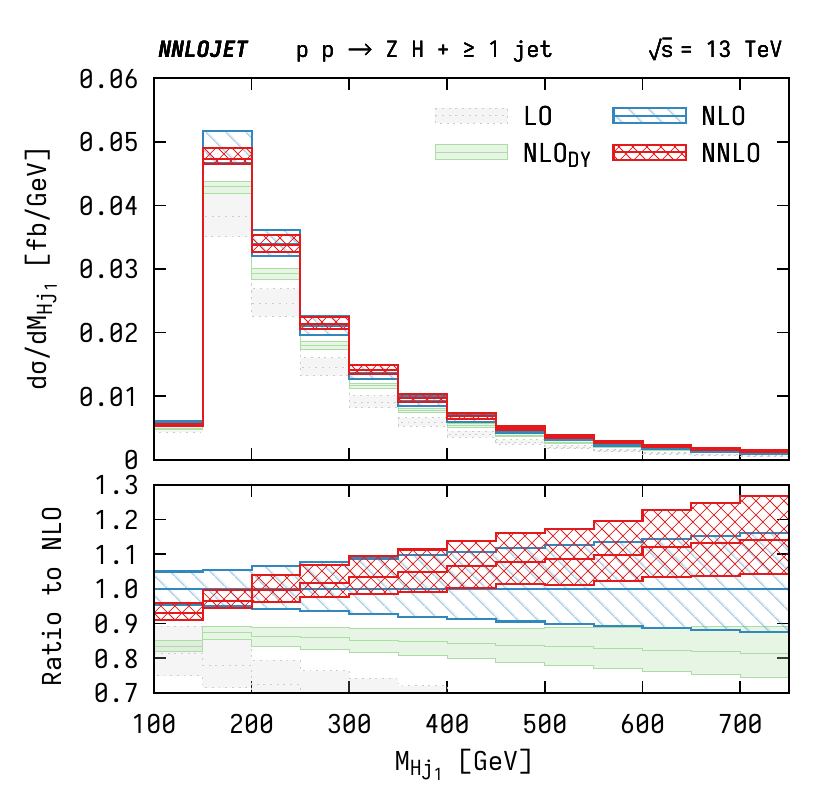}
    \subcaption{}
    \label{fig:z_mhj1}
  \end{subfigure}
  \begin{subfigure}[h]{.4\textwidth}
    \includegraphics[width=\textwidth]{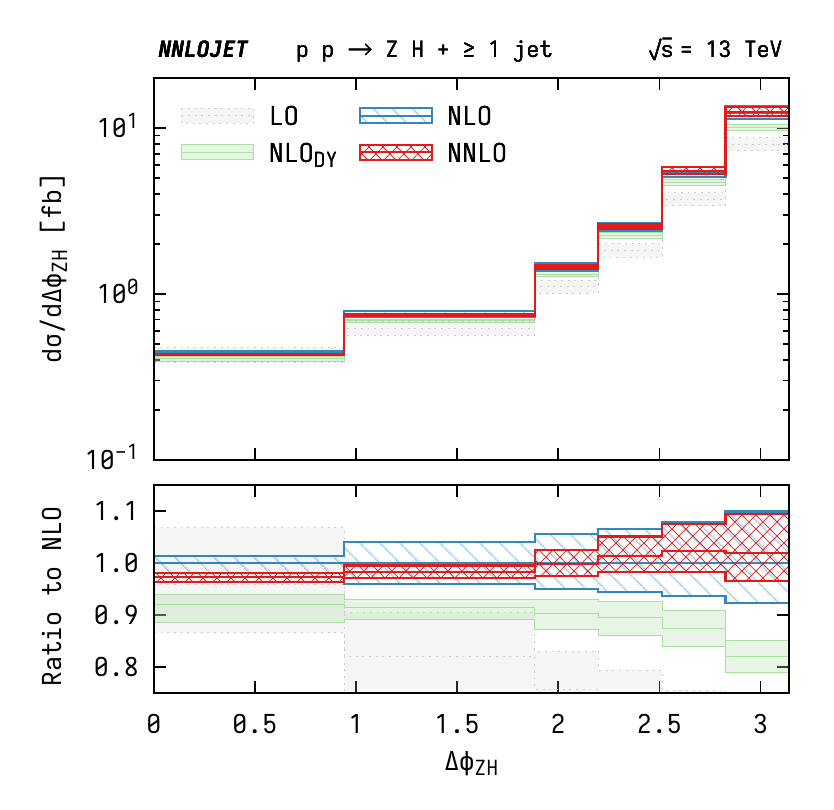}
    \subcaption{}
    \label{fig:z_deltaphi}
  \end{subfigure}
  \begin{subfigure}[h]{.4\textwidth}
    \includegraphics[width=\textwidth]{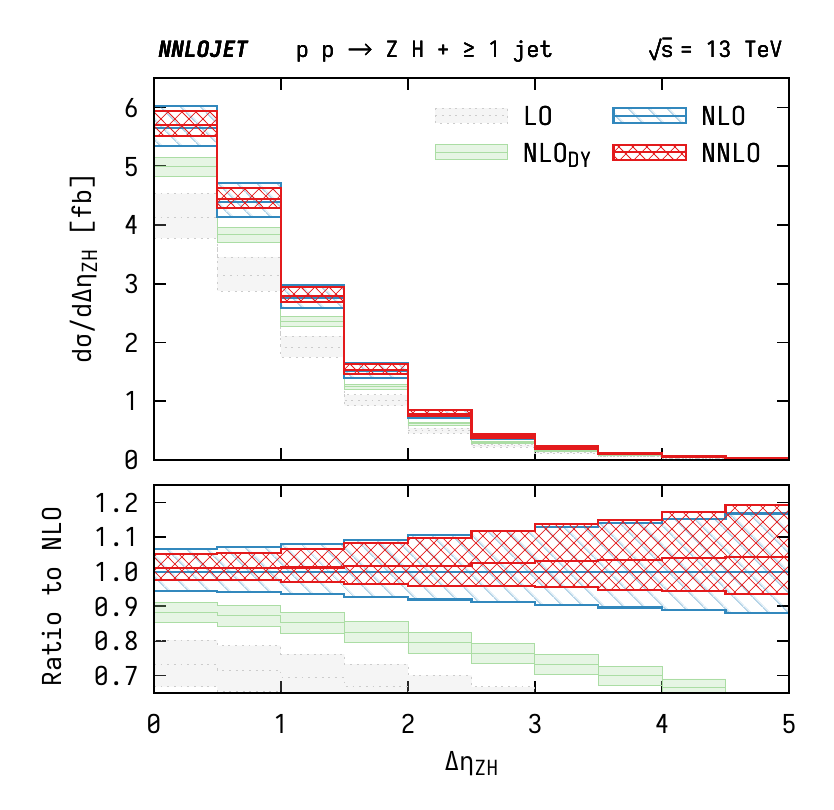}
    \subcaption{}
    \label{fig:z_deltaeta}
  \end{subfigure}
  \begin{subfigure}[h]{.4\textwidth}
    \includegraphics[width=\textwidth]{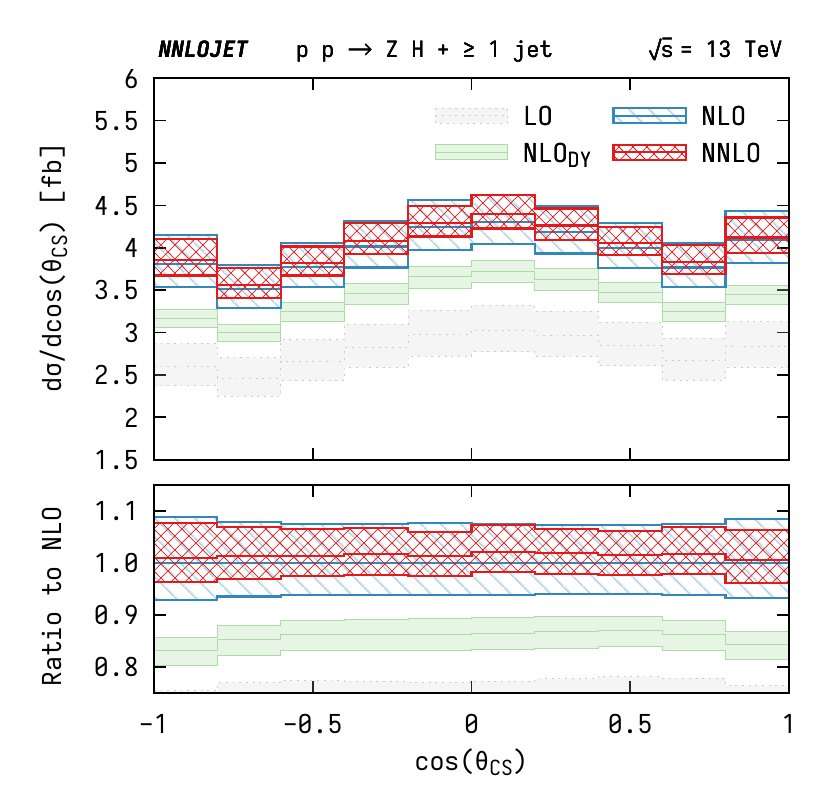}
    \subcaption{}
    \label{fig:z_costheta}
  \end{subfigure}
  \caption{ZH+J observables for ZH signal : (a) $M_{\PH j_1}$:  the  invariant mass distribution of the Higgs boson--jet system, 
  (b) $\Delta\phi_{\PZ\PH}$: the azimuthal angle separation \PZ and \PH boson, (c) $\Delta\eta_{\PZ\PH}$: the pseudo-rapidity separation between the \PZ and \PH boson, (d) $\cos(\theta_\text{CS})$: the Collins-Soper angle.}
  \label{fig:nc_more}
\end{figure}

Finally, we discuss results that are specific to the neutral-current Higgs boson production process. To this end, we consider observables that are motivated by the experimental measurements and used in multivariate discriminants to improve the sensitivity of the $Z\PH$ analyses~\cite{Aad:2020jym}.

Figure~\ref{fig:z_mhj1} shows the invariant mass distribution of the Higgs boson--jet system in $\PZ\PH+\jet$ inclusive production.
This observable is important for the rejection of multi-jet backgrounds.
For the production of a Higgs boson and a jet of $p_\rT^{j}>20~\GeV$ and 
above the kinematic threshold of $M_{\PH j_1}\sim 150~\GeV$, where a maximum is reached, the spectrum is steeply falling. 
Focussing on the second panel of this figure ~\ref{fig:z_mhj1}, we observe that the relative size of the NNLO corrections increases from $-10\%$ at $M_{\PH j_1} \sim 100$ GeV to $+30\%$ at $M_{\PH j_1} \sim 750$ GeV with  residual scale uncertainty bands increasing from $\pm 3$\% to $\pm 10\%$ over the same range.
This behaviour can be attributed to the fact that at high invariant mass, multi-jet configurations, included here at NLO level only, dominate compared to the exclusive 1-jet production process, which is here clearly suppressed.  

The angular separation between the Higgs boson and the \PZ boson is an important input to identify the Higgs boson candidate and can be probed via the azimuthal angle and the pseudo-rapidity separation variables shown in figures~\ref{fig:z_deltaphi} and \ref{fig:z_deltaeta}, respectively.

As argued in section~\ref{sec:inc_excl}, the \PZ and \PH bosons are preferably produced back-to-back in the transverse plane.
This statement is further corroborated by the azimuthal angle distribution shown in figure~\ref{fig:z_deltaphi} that strongly peaks around $\Delta\phi_{\PZ\PH}\sim\pi$.
While additional emissions further open up the phase space for $\Delta\phi_{\PZ\PH}\to0$, the region below this peak value still remains severely suppressed given the large mass of the two objects. For values of the azimuthal angle separation close to zero, the cross section value is also close to zero. This is clearly seen in the first panel of this figure~\ref{fig:z_deltaphi}.
To avoid unstable bin to bin fluctuations, for both of the panels in  this particular figure~\ref{fig:z_deltaphi} the size of the corrections in the first two bins has been averaged.
Looking at the second panel, we see here that, unlike in all the distributions discussed so far in this inclusive neutral-current mode, the size of the corrections is the largest where the cross section is the largest. 
Indeed, close to the peak value of $\Delta\phi_{\PZ\PH}\sim\pi$, attained exactly in a leading order back-to-back configuration for ZH production, the size of the NNLO corrections relative to NLO and the corresponding uncertainties 
are at the level of $10\%$ while they decrease to be at percent level for $\Delta\phi_{\PZ\PH}\to0$.

Figure~\ref{fig:z_deltaeta} shows the pseudo-rapidity separation between the \PZ and \PH boson. The distribution is peaked around zero, indicating that both particles are produced at central rapidities in the partonic collision frame. It then decreases steeply towards higher values of the pseudo-rapidity separation variable, yielding a null cross section at values close to 
$\Delta\eta_{\PZ\PH} \sim 5$. This is clearly seen by analysing the results presented in the first panel of figure ~\ref{fig:z_deltaeta}. 
Analysing the K-factors relative to NLO presented in the second panel, and comparing the (N)NLO predictions with the NLO$_{\DY}$ one, 
we further conclude that the pure \DY-type corrections are dominant at small to medium values of the pseudo-rapidity separation, whereas the quark-loop induced \GGF contributions have their most important impact at large rapidity separation where the cross section becomes small. In this high rapidity regime, the size of the corrections (and of the uncertainty bands) is enlarged and can reach values up to $20\%$.

In figure~\ref{fig:z_costheta} we present results for the cosine of the Collins-Soper angle, which is defined as the polar angle of the (negatively charged) lepton in the \PZ boson rest frame.
This observable is sensitive to the polarisation of the \PZ boson and helps in distinguishing the $\PZ\PH$ signal from the $\PZ+\jet$ backgrounds.
Being defined in the rest frame of the gauge boson and considering the fact that the leptonic decay $\PZ\to\Plm\Plp$ does not take part in any QCD interactions, the observable $\cos(\theta_\text{CS})$ is largely insensitive to the recoil from the jet and other QCD emissions.
As a consequence, QCD corrections to this observable lead to remarkably flat $K$-factors (close to one), that are essentially given by those of the fiducial cross sections given in table~\ref{zhfid}.
This is best seen comparing the results labelled as (N)NLO and NLO$_{\DY}$ in the second panel of figure~\ref{fig:z_costheta}. 

For the majority of the distributions presented here, we observe that at low or medium values of the kinematical variable probed and corresponding to a region where the differential cross sections are the largest, the impact of the NNLO corrections are mainly due to genuinely new Drell-Yan type contributions present at order $\alphas^3$.  
The inclusion of those corrections lead to a clear stabilisation of the predictions and a considerable reduction of the left over theoretical uncertainty band. 
The inclusion of the quark-loop induced contributions is relevant to achieve percent-level accuracy in those regions. Instead, at higher values of the kinematical variable probed, the impact of the quark-loop induced contributions becomes significantly more important
%
and one observes a magnification of the size of the corrections, a distortion of the distribution shape, and an enlargement of the left-over remaining uncertainties in the NNLO predictions.

\section{Conclusions and Outlook}
\label{conclusions}

In this work we have presented state-of-the-art predictions for the charged- and neutral-current $V\PH+\jet$ production processes.
These include the corrections to the Drell-Yan type contributions up to order $\alphas^3$, i.e.\ at NNLO in perturbative QCD, as well as the leading heavy quark loop induced contributions up to order $\alphas^3$.
These calculations were used to perform phenomenological studies of all three $V\PH+\jet$ production modes in proton--proton collisions at $\sqrt{s} = \SI{13}{\TeV}$.
In particular, a range of differential observables have been considered both for exclusive ($n_\jet = 1$) and inclusive ($n_\jet \geq 1$) jet production.

One of the motivations for considering $V\PH+\jet$ production is that event selection categories in the experimental analyses with $n_\jet \geq 1$ can often receive the largest contribution of signal events in an experimental set-up~\cite{ATLAS:2020fcp}.
Note that requiring a jet in association with the $V\PH$ final state typically lowers the perturbative accuracy of a fixed order in $\alphas$ prediction, which therefore increases the relative theoretical uncertainty associated with the modelling of the $V\PH+\jet$ signal in these categories (and overall reduces the sensitivity to interpret the Higgs data).
It is for these reasons that dedicated calculations applied to the $V\PH+\jet$ categories (such as those presented here), are vital for future measurements/interpretations in the $V\PH$ channel.

We have found that the inclusion of higher-order QCD corrections to the Drell-Yan type contributions are essential in stabilizing the predictions and in reducing the theoretical uncertainty for both charged- and neutral-current production modes (see also~\cite{Gauld:2020ced} where we presented selected results for $\PWp\PH+\jet$ production).  
This is particularly true in the kinematical regimes associated with low to medium values of the transverse momentum of the produced vector boson and where the differential cross sections are the largest.  

The heavy quark loop induced contributions to the charged-current processes of order $\alpha_s^2 y_t$ introduce corrections at the percent level. While knowledge of the QCD correction to these contributions is desirable, the (known) leading contributions of order $\alpha_s^2 y_t$ are likely sufficient to obtain cross-section predictions accurate at the percent-level.

For the neutral-current process, we find that the heavy quark loop induced contributions have their largest phenomenological impact via the gluon-gluon fusion type contributions, named as \GGF, and first present at order $\alphas^3$. 
We find that, due to the inclusion of these quark-loop induced contributions in the computations one observes an increase in the size of the NNLO corrections, a distortion of the distribution shape, and an enlargement of the left over remaining uncertainties in the NNLO predictions. In most cases, however, these effects appear enhanced in kinematical 
regions associated to large values of $p_{\rT,\PZ}$ (typically above $150~\GeV$) where the cross sections are smaller.
A reduction of this source of theoretical uncertainty will likely require the computation of order $\alphas^4$ corrections. The calculation of the necessary two-loop amplitudes enabling such a computation will be a challenging task.

Overall, the calculations presented in this work are critical for improving the precision of the Higgs boson signal templates~\cite{LHCHiggsCrossSectionWorkingGroup:2016ypw,Andersen:2016qtm}, which play an essential role in the measurement and interpretation of Higgs boson production data at the LHC.
In particular, these calculations are relevant for both the exclusive ($n_\jet = 1$) and inclusive ($n_\jet \geq 1$) event selections, for both charged- and neutral-current processes in the $V\PH$ production mode.

In the long term, the fixed-order calculations presented here also serve as an important milestone towards achieving an \NNLO\!+PS accurate~\cite{Mazzitelli:2020jio} description of the $V\PH+\jet$ process, deriving jet-vetoed cross sections at \N3\LO for associated Higgs production, and eventually aiming for a fully differential \N3\LO calculation for the $V\PH$ process.
Given the large data samples which are anticipated to be collected during the high-luminosity phase of the LHC, this level of theoretical precision is desirable to increase the sensitivity of future Higgs boson measurements.

\acknowledgments

We would like to thank Jonas Lindert for facilitating the use and inclusion of OpenLoops amplitudes into our computations, and to Hannah Arnold, Brian Moser, Tristan du Pree for discussions on experimental aspects of this work.
Furthermore we thank Xuan Chen, Juan Cruz-Martinez, James Currie, Thomas Gehrmann, Marius Höfer, Matteo Marcoli, Jonathan Mo,Tom Morgan, Jan Niehues, João Pires, Robin Sch\"urmann, Duncan Walker, and James Whitehead for useful discussions and their many contributions to the \nnlojet code.

This research is supported by the Dutch Organisation for Scientific Research (NWO) through the VENI grant 680-47-461, by the UK Science and Technology Facilities Council (STFC) through grant ST/T001011/1, by the Swiss National Science Foundation (SNF) under contract 200021-172478 and by the Swiss National Supercomputing Centre (CSCS) under project ID ETH5f.

\bibliography{VHJ}

\end{document}